\documentclass[a4paper,11pt]{article}
\usepackage{amsfonts,epic,graphicx}

\input{tcilatex}

\begin{document}

\setcounter{page}{0} \topmargin0pt \oddsidemargin5mm \renewcommand{%
\thefootnote}{\fnsymbol{footnote}} \newpage \setcounter{page}{0} 
\begin{titlepage}
\begin{flushright}
Berlin Sfb288 Preprint  \\
hep-th/0103055
\end{flushright}
\vspace{0.2cm}
\begin{center}
{\Large {\bf Scaling functions from q-deformed Virasoro characters} }

\vspace{0.8cm}
{\large  O.A.~Castro-Alvaredo  and  A.~Fring}

\vspace{0.2cm}
{Institut f\"ur Theoretische Physik, 
Freie Universit\"at Berlin,\\
Arnimallee 14, D-14195 Berlin, Germany }
\end{center}
\vspace{0.5cm}
 
\renewcommand{\thefootnote}{\arabic{footnote}}
\setcounter{footnote}{0}

\begin{abstract}
We propose a renormalization group scaling function which is constructed from
q-deformed  fermionic versions of Virasoro characters. By comparison
with alternative methods, which  take their starting point  in the massive theories, we
demonstrate that these new functions contain qualitatively the same information. We
show that these functions allow for RG-flows not only amongst members of a particular 
series of conformal  field theories, but also between different series such as $N=0,1,2$
supersymmetric conformal field theories.  We provide a
detailed analysis of how Weyl characters may be utilized in order to solve various 
recurrence relations emerging at the fixed points of these flows.  
The q-deformed Virasoro characters allow
furthermore for the construction of particle spectra, which involve unstable pseudo-particles.
\end{abstract}
\vfill{ \hspace*{-9mm}
\begin{tabular}{l}
\rule{6 cm}{0.05 mm}\\
Olalla@physik.fu-berlin.de\\
Fring@physik.fu-berlin.de
\end{tabular}}
\end{titlepage}
\newpage

\section{Introduction}

Renormalization group (RG) methods have been developed \cite{GL} to carry
out qualitative studies of regions of quantum field theories which are not
accessible to perturbation theory in the coupling constant. For theories in
1+1 space-time dimensions these methods admit particularly powerful
realizations in form of explicit constructions of scaling functions. Such
functions may be obtained either from the thermodynamic Bethe ansatz (TBA) 
\cite{TBA}, from correlations functions involving various components of the
energy-momentum tensor \cite{ZamC,CF2} or from semi-classical studies \cite
{Zamref}. In general the functions obtained from different approaches differ
quantitatively, but nonetheless possess the same qualitative features
characterized as follows:

We consider a quantum field theory which contains asymptotically stable
particles of mass $m_{i}$ and unstable particles with mass $M_{i}$. In
addition we assume that there are no particles associated to asymptotic
massless states in the spectrum. Then the scaling function $c(r)$
parameterized by a dimensionless renormalization group parameter $r\,$ has
the properties: It

\begin{enumerate}
\item[i)]  coincides with the Virasoro central charge $c$ of the ultraviolet
conformal field theory for vanishing $r$%
\begin{equation}
\lim_{r\rightarrow 0}c(r)=c,
\end{equation}

\item[ii)]  is non-increasing along the RG flow,

\item[iii)]  is stationary at RG fixed points and acquires at these points
the Virasoro central charge of specific conformal field theories 
\begin{equation}
c(r)=c_{ij}=const\qquad m_{i},M_{i}\ll \frac{2}{r}\ll m_{j},M_{j}\,,
\end{equation}

\item[iv)]  vanishes in the infrared 
\begin{equation}
c(r)=0\qquad \frac{2}{r}\ll m_{i},M_{i}\,\,.
\end{equation}
\end{enumerate}

There is yet another proposal to construct such type of functions, namely as
``Bailey flow'' \cite{Foda} between different series of Virasoro characters.
However, so far it has neither been established whether the functions
constructed in this fashion satisfy the properties i)-iv) nor has it been
clarified in which way they are related to a massive quantum field theory.

In the following we shall be constructing a scaling function which also
flows between certain Virasoro characters. In addition to the flows provided
in \cite{Foda}, we will not only propose a flow between several distinct
series, such as for instance from N=2 superconformal theories to N=1
superconformal theories, but also realize the flows within a particular
series itself. Our flows are manifested by means of q-deformed Cartan
matrices which simulate a control of the energy scales of unstable
particles. We establish that the proposed function indeed satisfies the
properties i)-iv) and in addition relate it to a concrete massive quantum
field theory with an explicitly known scattering matrix.

Our manuscript is organized as follows: In section 2 we recall how certain
recurrence relations emerge from a saddle point analysis of fermionic
versions of Virasoro characters, which involve data of the massive theory,
namely the phase of the scattering matrix, and how their solutions are
related to the effective central charge. We show that various series may be
realized in terms of the HSG-models. In section 3 we present a q-deformed
version of the analysis in section 2 and demonstrate how the HSG-realization
allows for a flow amongst various models governed by the mass scales of the
unstable particles. The analysis in this section is mainly carried out
numerically. Section 4 is devoted to the explicit analytic solutions at the
plateaux in terms of Weyl characters. We present here various cases which
have not been considered before. In section 5 we demonstrate how the
q-deformed characters may be associated to particle spectra, which involve
also unstable pseudo-particles. Our conclusions are stated in section 6.

\section{The TBA from the massive and massless side}

Let us first recall some well-known facts in order to assemble the relevant
equations and to establish our notations. We consider a Virasoro character
in the so-called ``fermionic version''\footnote{%
In fact this terminology is slightly misleading, since they are not
intrinsically fermionic. This name originated from the construction of
fermionic pseudo-particle spectra. However, it is also possible to construct
from (\ref{chi}) pseudo-particle spectra related to all kinds of general
statistics.}\cite{KM} 
\begin{equation}
\chi (q)=\sum\limits_{\vec{m}\in \frak{S}}^{\infty }q^{\vec{m}M\,\vec{m}%
^{t}/2+\vec{m}\cdot \vec{B}}\,\prod_{i=1}^{l}\,\left[ \QATOPD. . {(\vec{m}%
(1-M))_{i}+B_{i}^{\prime }}{m_{i}}\right] _{q}.  \label{chi}
\end{equation}
Here we employ the standard abbreviation for Euler's function $(q)_{m}$ with 
$(q)_{0}=1$ and the Gau\ss ian polynomial (q-binomial), see e.g. \cite{Andr}%
, for the integers $n$ and $m$ with $0\leq m\leq n$%
\begin{equation}
(q)_{m}:=\prod_{k=1}^{m}(1-q^{k})\quad \quad \text{and\qquad }\,\left[
\QATOPD. . {n}{m}\right] _{q}:=\frac{(q)_{n}}{(q)_{m}(q)_{n-m}}\,\,.
\end{equation}
The main characteristics of the expression (\ref{chi}) for the character $%
\chi (q)$ are the real symmetric ($l\times l$)-matrix $M$ and the vector $%
\vec{B}^{\prime }$ with $B_{i}^{\prime }=\infty $ for $1\leq i\leq
l-l^{\prime }$, $B_{i}^{\prime }=0$ for $l-l^{\prime }<i\leq l$, with $%
l^{\prime }$ being a non-negative integer smaller $l$. The specific form of
the vector $\vec{B}$ distinguishes between different highest weight
representations, which share of course the same Virasoro central charge $c$.
There might be restrictions on the set $\frak{S}$ in which $\vec{m}$ takes
its values, which usually reflect some of the symmetries in the model.

The important thing for us to note is that once $\chi (q)$ is of the generic
form (\ref{chi}), one may employ the techniques originally pursuit in \cite
{Rich} and carry out a saddle point analysis to extract the leading order
behaviour. As a result of this, the effective central charge, i.e. $c_{\text{%
eff}}=c-24h^{\prime }$ with $h^{\prime }\,$ being the smallest conformal
dimension occurring in the theory ($h^{\prime }=0$ in unitary models), is
expressed in a rather non-obvious way. For the character of the particular
form (\ref{chi}), this analysis was performed first in \cite{KM}, leading,
after a suitable variable transformation, to the saddle point conditions 
\begin{equation}
1-x_{A}=\prod_{B=1}^{l}(x_{B})^{M_{AB}}\qquad \text{and\qquad }%
1-y_{A}=\prod_{B=1+l-l^{\prime }}^{l}(y_{B})^{M_{AB}^{\prime }}\,\,.
\label{cTBA}
\end{equation}
At this stage the $x_{A}$ and $y_{A}$ are just the integration variables
occurring in this context (for details see e.g. \cite{Rich,KM}). The matrix $%
M^{\prime }$ is a submatrix of $M$ of dimension ($l^{\prime }\times
l^{\prime }$). The remaining $y$'s which do not occur in these equations are
taken to be one, i.e. $y_{A}=1$ for $1\leq A\leq l-l^{\prime }$. One should
also note that, since in this analysis sums are converted into integrals,
the specific structure of the set $\frak{S}$ \ does not effect the outcome
of the computation and may therefore be ignored for our purposes. The
leading order behaviour at the extremum point yields the effective central
charge 
\begin{equation}
c_{\text{eff}}=\frac{6}{\pi ^{2}}\sum\limits_{A=1}^{l}\left( \mathcal{L}%
\left( 1-x_{A}\right) -\mathcal{L}\left( 1-y_{A}\right) \right) \,\,
\label{ceff}
\end{equation}
in terms of Rogers dilogarithm $\mathcal{L}(x)=\sum_{n=1}^{\infty
}x^{n}/n^{2}+\ln x\ln (1-x)/2$ (for properties see e.g. \cite{Log}). Once $%
c_{\text{eff}}$ is rational, the system (\ref{cTBA}) and (\ref{ceff}) is
referred to as ``accessible'' dilogarithms (for a review see e.g. \cite
{dilog} and references therein), which from the mathematical point of view
is a rather exceptional situation.

The important point to note here is that the saddle point analysis does not
rely upon the fact that the matrices $M$ and $M^{\prime }$ are constant. It
is this feature which we shall exploit below.

\subsection{\textbf{g\TEXTsymbol{\vert}\~{g}}-theories}

Intriguingly the same set of equations (\ref{cTBA}) and (\ref{ceff}) may
also be obtained when we commence with the massive instead of the conformal
side. We start from a scattering matrix $S_{AB}(\theta )$, as a function of
the rapidity $\theta $, between particles of type $1\leq A,B\leq l$.
Performing then a thermodynamic Bethe ansatz analysis \cite{TBA} one ends up
with a set of non-linear integral equations in the pseudo-energies as
functions of the rapidities, the so-called TBA-equations. We then assume
that the S-matrix is such that it leads to regions in the TBA-equations in
which the pseudo-energies are constant. In general this happens when the
scattering matrix does not depend on the effective coupling constant. In
that situation, the thermodynamic Bethe Ansatz leads to a set of coupled
equations coinciding precisely with the ones in $x$ in (\ref{cTBA}). All $y$%
's may be thought of as being 1 in this case. The matrix $M$ in (\ref{chi})
is now directly related to the massive models containing the information
about the scattering matrix 
\begin{equation}
M_{AB}=\delta _{AB}-\frac{1}{2\pi i}\lim_{\theta \rightarrow \infty }\ln
(S_{AB}(\theta )S_{BA}(\theta ))\,\,.  \label{ph}
\end{equation}
Reversing the argument, the relation (\ref{ph}) means that one has
identified a quantity within the conformal field theory which carries the
data of the phase of the S-matrix.

In the following we will consider theories in which $M_{AB}$ is related to a
Lie algebraic structure. For this purpose we give the quantum numbers $A,B$,
which describe the particle type, an additional substructure. We identify
each particle by two quantum numbers, i.e. $A=(a,i)$, such that the
scattering matrices are of the general form $S_{ab}^{ij}(\theta )$. We
associate the main quantum numbers $a,b$ to the vertices of the Dynkin
diagram of a simply laced Lie algebra \textbf{g }of rank $\ell $ and the
so-called colour quantum numbers $i,j$ to the vertices of the Dynkin diagram
of a simply laced Lie algebra \textbf{\~{g}} of rank $\tilde{\ell}$. We
refer to these theories as \textbf{g\TEXTsymbol{\vert}\~{g}}. The S-matrices
constructed in \cite{FK} are of this type 
\begin{equation}
S_{ab}^{ij}(\theta )=e^{i\pi \varepsilon _{ij}K_{\bar{a}b}^{-1}}\exp
\int\limits_{-\infty }^{\infty }\frac{dt}{t}\left( 2\cosh \frac{\pi t}{h}-%
\tilde{I}\right) _{ij}\left( 2\cosh \frac{\pi t}{h}-I\right)
_{ab}^{-1}e^{-it(\theta +\sigma _{ij})},  \label{SCO}
\end{equation}
with $I,\tilde{I}$ being the incidence matrix of \textbf{g, \~{g}, }%
respectively. Here $\varepsilon _{ij}$ is the Levi-Civita pseudotensor, $h$
the Coxeter number of \textbf{g }and $\sigma _{ij}=-\sigma _{ij}$ the
resonance parameters. As special cases of this S-matrix we have the \textbf{g%
\TEXTsymbol{\vert}A}$_{1}$ and \textbf{A}$_{n}$\textbf{\TEXTsymbol{\vert}%
\~{g} }theories which correspond to the minimal affine Toda theories and the 
\textbf{\~{g}}$_{n+1}$-homogeneous Sine-Gordon (HSG) models \cite{HSG}. As
may be seen \cite{FK} easily from (\ref{SCO}) the M-matrix for these models
is 
\begin{equation}
M_{ab}^{ij}=\,K_{ab}^{-1}\tilde{K}_{ij}\,,  \label{M}
\end{equation}
with $K,\tilde{K}$ being the Cartan matrix of \textbf{g, \~{g}, }%
respectively. The special case \textbf{g\TEXTsymbol{\vert}A}$_{1}$ was first
treated in \cite{TBAKM}. S-matrices for \textbf{\~{g} }also to be non-simply
laced were proposed in \cite{CK}. It remains an open question, apart from 
\textbf{g\TEXTsymbol{\vert}A}$_{1}$, how to allow also \textbf{g} to be
non-simply laced.

\subsection{\textbf{g\TEXTsymbol{\vert}\~{g}}--coset theories}

The full system (\ref{cTBA}) and (\ref{ceff}), involving a non-trivial $%
M^{\prime }$-matrix, can be associated in general with a non-diagonal
scattering matrix on the massive side. A straightforward identification
between $M$ and the scattering matrix such as in (\ref{ph}) is not possible
in this case. However, within the thermodynamic Bethe ansatz analysis the
equations are diagonalized and decoupled, such that at the fixed points they
acquire precisely the form (\ref{cTBA}). In many prominent cases the $M$ and 
$M^{\prime }$ matrices involve Lie algebraic quantities in the form of (\ref
{M}). Noting this point, many models can be realized formally in terms of 
\textbf{g\TEXTsymbol{\vert}\~{g}}-cosets.\ 

\subsubsection{Unitary minimal models}

The series of unitary minimal models, usually denoted by $\mathcal{M}(k,k+1)$
\cite{BPZ}, constitute an extremely well studied and prominent class of
conformal field theories. It is well-known \cite{GKO} that they may for
instance be realized by the cosets $SU(2)_{k}\otimes U(1)/SU(2)_{k+1}$ or $%
SU(k+1)_{2}/SU(k)_{2}\otimes U(1)$, which are related to each other by
level-rank duality \cite{LR}. Recalling the fact \cite{GKO} that each
extended simple Lie algebra $g$, a Kac-Moody algebra $\hat{g}$ of level $k$,
contributes positively or negatively $k\dim g/(k+h)$ ($h$ being the Coxeter
number of $g$) to the total central charge, depending on whether it is part
of the algebra or subalgebra, respectively, one obtains the famous sequence 
\begin{equation}
c=1-\frac{6}{(k+2)(k+3)}\,\qquad k=1,2,3,\ldots  \label{N0}
\end{equation}
Including now the relevant $U(1)$-factors, we may also obtain the series (%
\ref{N0}) from a coset of two \textbf{g\TEXTsymbol{\vert}\~{g}}--theories 
\begin{equation}
A_{k-1}|A_{1}/A_{k}|A_{1}\qquad \Leftrightarrow \qquad
A_{1}|A_{k}/A_{1}|A_{k-1}\qquad  \label{HSGmin}
\end{equation}
in the ultraviolet limit. The relation (\ref{HSGmin}) allows for various
interpretations with regard to the realizations of several RG-flows. We note
that both theories on the l.h.s. do not contain any unstable particle. A
flow between cosets parameterized by different $k$'s may then be achieved in
the so-called massless way as roaming trajectories in the spirit of \cite
{Stair}. On the other hand, the realizations in form of the r.h.s. of (\ref
{HSGmin}) constitute theories which contain unstable particles. Therefore a
flow between cosets related to different $k$'s is achievable in a well
controllable fashion over the different energy scales of the unstable
particles as observed in \cite{CFKM,CFK,CF1,CF2,CF3} for the HSG-models. For
vanishing resonance parameters $\sigma _{ij}$ the system on the r.h.s. of (%
\ref{HSGmin}) leads to the same constant TBA-equations as found for the
RSOS-models \cite{RSOS}. In addition following the RG-flow of the scaling
function of the TBA one observes that at the fixed points, the set of
equations (\ref{cTBA}) is also obtained for finite values of the resonance
parameters.

Of course these coset realizations are not unique and one may for instance
also obtain (\ref{N0}) from the quaternionic projective space $HP^{k}$ \cite
{GKO} or use various exceptional Lie algebras to construct particular
theories. This ambiguity allows for various other realizations in terms of
different combinations of HSG-models.

\subsubsection{Unitary N=1 super conformal field theories}

The series of $N=1$ unitary minimal models $\mathcal{M}^{N=1}(k,k+1)$ has
played an important role in the construction of certain string theories. It
may be realized for instance by the cosets $SU(2)_{k}\otimes
SU(2)_{2}/SU(2)_{k+1}$ or $SU(k+2)_{2}/SU(k)_{2}\otimes SU(2)_{2}$ \cite{GKO}%
. The corresponding series for the Virasoro central charge is 
\begin{equation}
c=\frac{3}{2}-\frac{12}{(k+2)(k+4)}\,\qquad k=1,2,3,\ldots  \label{N1}
\end{equation}
Once again we may include the relevant $U(1)$-factors and also construct the 
$\mathcal{M}^{N=1}(k,k+1)$ models from several \textbf{g\TEXTsymbol{\vert}%
\~{g}}--theories, for instance 
\begin{equation}
A_{k-1}|A_{1}\otimes A_{1}|A_{1}\otimes A_{1}|A_{1}/A_{k+1}|A_{1}\qquad
\Leftrightarrow \qquad A_{1}|A_{k+1}/A_{1}|A_{k-1}\otimes
A_{1}|A_{1}\,\,.\qquad  \label{HSGN1}
\end{equation}
In the ultraviolet limit they posses central charges of the form (\ref{N1}).
Once again we note that there is a realization which involves unstable
particles, i.e. the r.h.s. of (\ref{HSGN1}), and one which does not, that is
the l.h.s. of (\ref{HSGN1}).

\subsubsection{Unitary N=2 super conformal field theories}

The series of $N=2$ unitary minimal models $\mathcal{M}^{N=2}(k,k+1)$ is
omnipresent in string theory \cite{N2} (for a recent review see e.g. \cite
{Gato}). It may be realized by the cosets $SU(2)_{k}\otimes U(1)/U(1)_k$ 
or $SO(2k)_{2}/SU(k)_{2}$ with the corresponding series of the 
Virasoro central
charge 
\begin{equation}
c=\frac{3k}{2+k}\,\qquad k=1,2,3,\ldots  \label{N2}
\end{equation}
Including the relevant $U(1)$-factors, we construct from several \textbf{g%
\TEXTsymbol{\vert}\~{g}}--theories the realizations 
\begin{equation}
A_{k-1}|A_{1}\otimes A_{3}|A_{1}\qquad \Leftrightarrow \qquad
A_{1}|D_{k+1}/A_{1}|A_{k-1}\qquad .  \label{HSGN2}
\end{equation}
In the ultraviolet limit they also lead to (\ref{N2}). A further
possibility, which we shall exploit in section 3.4., to obtain (\ref{N2}),
is to use the coset $A_{1}|D_{k+2}/A_{1}|A_{k-1}\otimes A_{1}|A_{1}^{\otimes
2}$. Once again we note that there is a realization which involves unstable
particles, i.e. the r.h.s. of (\ref{HSGN2}), and one which does not, that is
the l.h.s. of (\ref{HSGN2}).

\subsubsection{$G_{k}\otimes G_{l}/G_{k+l}$-cosets}

The $G_{k}\otimes G_{l}/G_{k+l}$-cosets are more general theories which
encompass various models. For instance taking \ $G=SU(2)$ and setting $l=2$
or $l=k-2,\,k=1$ one obtains the $\mathcal{M}^{N=1}(k,k+1)$ or $\mathcal{M}%
(k,k+1)$-models, respectively. Massless flows related to these models where
investigated in \cite{Stair}. Once again there exists a realization in terms
of HSG-models 
\begin{equation}
A_{k-1}|G\otimes A_{l-1}|G\otimes A_{1}|A_{1}^{\otimes 2\ell }/A_{k+l-1}|G,
\label{pp}
\end{equation}
such that we may also reproduce these flows by means of a variation of the
energy scales of the unstable particles. Here $\ell $ is still the rank of
the Lie algebra g. We will not perform a detailed investigation of these
theories which go beyond the $\mathcal{M}^{N=1}(k,k+1)$ or $\mathcal{M}%
(k,k+1)$-models, but from the following analysis it will become apparent
that the existence of the realization (\ref{pp}) allows for an analogue
treatment.

\section{RG-flow from q-deformed Virasoro characters}

We now wish to introduce a mass scale. Recalling \cite{Resh,Kun} that the
recurrence relations (\ref{cTBA}) may be solved by means of Weyl characters
a natural conjecture is to suspect that a deformation of these expressions
leads to a correct description of the massive theories in the sense of the
full TBA-equations. To make this concrete seems a rather difficult task and
we therefore construct a scaling function in a different way, but
nonetheless in the spirit of the renormalization group ideas. Instead of
using a different parameterization for the Weyl characters, we deform the
Virasoro characters (\ref{chi}) in a very natural way. As was already
pointed out in the previous section, the saddle point analysis which leads
to the equations (\ref{cTBA}) and (\ref{ceff}), does not depend on the fact
whether the matrix $M$ is constant or variable. We can exploit this by
introducing mass scales in a rather suggestive fashion. Restricting
ourselves to the large class of simply laced \textbf{g\TEXTsymbol{\vert}\~{g}%
}-theories and cosets constructed from these theories as in section 2.2, we
replace now the $M$-matrix by a q-deformed version 
\begin{equation}
\left[ M_{ab}^{ij}\right] _{q}:=\,\left[ K_{ab}\right] _{q}^{-1}[\tilde{K}%
_{ij}]_{\tilde{q}_{ij}}\,\,,
\end{equation}
with 
\begin{eqnarray}
\left[ K_{ab}\right] _{q} &:&=K_{ab}q=\alpha _{a}\cdot \alpha
_{b}\,q\,=\alpha _{a}\cdot \alpha _{b}\,\exp (-mr/2\,)\,  \label{Kd1} \\
\lbrack \tilde{K}_{ij}]_{\tilde{q}_{ij}} &:&=2\delta _{ij}-[\tilde{I}_{ij}]_{%
\tilde{q}_{ij}}=\tilde{\alpha}_{i}\cdot \tilde{\alpha}_{j}\,\tilde{q}_{ij}=%
\tilde{\alpha}_{i}\cdot \tilde{\alpha}_{j}\exp (-mr/2\,(1-\delta
_{ij})e^{|\sigma _{ij}|/2})\,\,.  \label{Kd2}
\end{eqnarray}
Here the $\alpha _{i},\tilde{\alpha}_{i}$ are the simple roots of \textbf{g,
\~{g}}, respectively. In other words we re-defined the usual scalar product
between the simple roots or equivalently q-deformed the roots themselves.
The bracket $[\,\,]_{q}$ is not to be confused with the usual notation of
q-deformed integers. Q-deformations of a different nature have recently
played an important role in the context of the formulation of consistent
expressions for scattering matrices of affine Toda field theories related to
non simply laced Lie algebras \cite{FKS}. For the case at hand the
q-deformation is mainly inspired by the physics of the unstable particles.
The natural mass scale of the unstable particle $m_{\tilde{c}}\sim
mr/2\,e^{|\sigma _{ij}|/2}$, with $\sigma _{ij}$ playing the role of a
resonance parameter and $m$ of an overall mass scale, is introduced in $%
\tilde{K}$ in such a way that for $\sigma _{ij}\rightarrow \infty $, the
Cartan matrix of \textbf{\~{g} }decouples according to the ``cutting rule''
analyzed in \cite{CF3}. Notice that for $mr/2\,e^{\sigma _{ij}/2}\ll 1$ we
have $[\tilde{K}_{ij}]_{\tilde{q}_{ij}}\approx \tilde{K}_{ij}$, such that
the decoupling takes place at the same scale as in the massive models (see
e.g. equation (51) in \cite{CFKM} and also \cite{CF2,CF3}). In addition we
would like the particles to be massless in the infrared. Recalling that the
masses of the affine Toda field theories can be organized in form of the
Perron-Frobenius vector of the Cartan matrix, the deformation (\ref{Kd1})
achieves this goal. In the limit $r\rightarrow 0$ we recover the usual
Cartan matrix.

Of course the deformations of the type (\ref{Kd1}) and (\ref{Kd2}) are not
unique and one could try to find different realizations in order to
construct scaling functions. However, from the arguments just outlined they
appear to be the most natural ones.

\subsection{\textbf{g\TEXTsymbol{\vert}\~{g}}-theories}

Equipped with the matrices (\ref{Kd1}) and (\ref{Kd2}), the q-deformed
version of (\ref{chi}) acquires the form 
\begin{equation}
\chi (q,r,\vec{m},\vec{\sigma})=\sum\limits_{\vec{k}=0}^{\infty }\frac{q^{%
\frac{1}{2}\vec{k}[M]_{\{r,\vec{m},\vec{\sigma}\}}\vec{k}^{t}+\vec{k}\cdot 
\vec{B}}}{(q)_{k_{1}}\ldots (q)_{k_{n}}}\,\,.  \label{qVir}
\end{equation}
For simplicity we took here $l^{\prime }$ to be zero. We collect the $\tilde{%
\ell}-1$ linearly independent resonance parameters in the vector $\vec{\sigma%
}$ and the $\ell $ independent mass scales in $\vec{m}$. The RG scaling
parameter is denoted by $r$. To obtain the recurrence relations in a more
symmetric way it is convenient to introduce the variables $%
x_{a}^{i}=\prod_{b=1}^{\ell }(Q_{b}^{i})^{-K_{ab}}$. In terms of the
q-deformed analogues to these variables, $[x_{a}^{i}]_{q}=\prod_{b=1}^{\ell
}(Q_{b}^{i})^{-\left[ K_{ab}\right] _{q}}$, the saddle point analysis of (%
\ref{qVir}) leads to 
\begin{equation}
\prod_{b=1}^{\ell }Q_{b}^{i}(r,\vec{m},\vec{\sigma})^{-\left[ K_{ab}\right]
_{q}}+\prod_{j=1}^{\tilde{\ell}}Q_{a}^{j}(r,\vec{m},\vec{\sigma})^{-[\tilde{K%
}_{ij}]_{\tilde{q}_{ij}}}\,\,=1  \label{Qr}
\end{equation}
together with the associated scaling function 
\begin{equation}
c^{\mathbf{g}|\mathbf{\tilde{g}}}(r,\vec{m},\vec{\sigma})=\frac{6}{\pi ^{2}}%
\sum\limits_{a=1}^{\ell }\sum\limits_{i=1}^{\tilde{\ell}}\mathcal{L}\left(
\prod_{j=1}^{\tilde{\ell}}Q_{a}^{j}(r,\vec{m},\vec{\sigma})^{-\left[ \tilde{K%
}_{ij}\right] _{q_{ij}}}\right) \medskip \,\,.  \label{cqr}
\end{equation}
The recurrence relations (\ref{Qr}) play now an analogous role to the
TBA-equations. In order to make our main point, namely that (\ref{cqr})
indeed constitutes a scaling function which reproduces the characteristic
features of the theory, like the ones obtainable from the conventional TBA,
the scaled version of the c-theorem or a semi-classical analysis, we have to
establish that $c^{\mathbf{g}|\mathbf{\tilde{g}}}(r,\vec{m},\vec{\sigma})$
satisfies indeed the properties i)-iv) in the introduction.

Most straightforward to prove are the properties related to the extremal
limits. Property i) is easily established since by construction $c^{\mathbf{g%
}|\mathbf{\tilde{g}}}(0,\vec{m},\vec{\sigma})$ is the ultraviolet Virasoro
central charge. Property iv) follows from the following argument: Let us
first assume in (\ref{Qr}) that the $Q_{a}^{i}$'s are finite for $%
r\rightarrow \infty $. Taking then this limit leads to \ $%
1+(Q_{a}^{i})^{-2}=1$, such that our initial assumption can not hold and we
deduce that $\lim_{r\rightarrow \infty }$ $Q_{a}^{i}\sim \infty $. When we
want to avoid that the scaling function (\ref{cqr}) becomes complex we have
to assume that the $Q$'s are real. Additional support for this assumption
will be provided below just based on the structure of (\ref{Qr}) and a
possible physical interpretation. Thus taking now $Q\in \Bbb{R}$ each term
on the l.h.s. of (\ref{Qr}) has to be smaller than $1$, such that we deduce
for the infrared asymptotics of the first term 
\begin{equation}
\lim_{r\rightarrow \infty }e^{-mr/2}\sum_{b}K_{ab}\ln Q_{b}^{i}=0\,\,.
\label{as}
\end{equation}
Excluding the exotic case $\sum_{b}K_{ab}\ln Q_{b}^{i}=0$, we demand the
behaviour (\ref{as}) for each term in the sum and conclude that the second
term in (\ref{Qr}) is zero such that with $\mathcal{L}(0)=0$ we finally
conclude that property iv) holds.

The other properties are less straightforward to prove in complete
generality and we will be content to establish them on the base of explicit
case-by-case examples.

\subsection{$\mathbf{A}_{1}\mathbf{|\tilde{g}}\equiv \mathbf{\tilde{g}}_{2}$%
-HSG}

The $\mathbf{A}_{1}\mathbf{|\tilde{g}}$-theories are good theories to start
with, since they do not involve any stable particle fusing structure. In
addition several scaling functions have been obtained by a TBA analysis \cite
{CFKM} and also from the scaled version of the c-theorem \cite{CF2,CF3},
such that we have already data available to compare with. The equations (\ref
{Qr}) become in this case simply 
\begin{equation}
Q^{i}(r,m,\vec{\sigma})^{2}=Q^{i}(r,m,\vec{\sigma})^{2-2q}+\prod_{j=1}^{%
\tilde{\ell}}Q^{j}{}(r,m,\vec{\sigma})^{\left[ \tilde{I}_{ij}\right]
_{q_{ij}}}.  \label{aa}
\end{equation}
It is useful to treat the case $\mathbf{\tilde{g}=A}_{1}$ separately, since
it corresponds to the free fermion.

\subsubsection{The free fermion}

The free fermion is analytically solvable in several approaches and is
therefore an ideal example to illustrate that the various scaling functions
are quantitatively different but contain qualitatively the same information.
Equation (\ref{aa}) becomes in this case simply $Q^{2}=Q^{2-2q}+1$. It is
not possible to solve this relation analytically, but near the ultraviolet
we may approximate $q\approx 1$ such that its solution becomes $Q\sim \sqrt{2%
}$ for $rm/2\ll 1$, and therefore 
\begin{equation}
c^{\mathbf{A}_{1}\mathbf{|A}_{1}}(rm)\sim \frac{6}{\pi ^{2}}\mathcal{L}%
\left( 1/2\right) =\frac{1}{2}\,\ \qquad \text{for }rm/2\ll 1.  \label{LM991}
\end{equation}
We can compare this with the scaling function obtained as exact solution of
the full TBA analysis 
\begin{equation}
c^{\text{TBA}}(rm)=\frac{6rm}{\pi ^{2}}\sum_{n=1}^{\infty }(-1)^{n}\frac{%
K_{1}(nrm)}{n}\sim \frac{1}{2}\,\ \qquad \text{for }rm/2\ll 1,  \label{LM992}
\end{equation}
where $K_{1}$ is a modified Bessel function. The latter estimate follows
from $K_{1}(rm)\sim 1/rm$ for $rm/2\ll 1$ and the fact that $\mathcal{L}%
\left( -1\right) =-12/\pi ^{2}$. This means in the main region of interest
these two functions coincide. It is also clear that for large $rm$ that both
functions vanish.

In addition we may compare with the scaling function obtained from the
c-theorem 
\begin{equation}
c^{\text{c-th}}(rm)=\frac{3}{2}\int_{rm}^{\infty }ds\,s^{3}\left(
K_{1}(s)^{2}-K_{0}(s)^{2}\right) \sim \frac{1}{2}\,\ \qquad \text{for }%
rm/2\ll 1  \label{LM993}
\end{equation}
which shows a similar behaviour. Note that despite the fact that we use $rm$
in (\ref{LM991})-(\ref{LM993}) the meaning of this parameter is different in
each context. For our purposes it is simply a dimensionless variable.

Let us now establish property ii) for this case. This illustrates at the
same time the general procedure which works in principle for all other
situations. Since we know that $Q(r=0)=\sqrt{2}$ and $\lim_{r\rightarrow
\infty }Q\rightarrow \infty $ we just have to establish that $Q(r)$ does not
posses a minimum or maximum in order to establish its monotonic behaviour.
We compute from (\ref{aa}) the derivative $Q^{\prime }=q\ln
Q/(2Q^{2q-1}-Q^{-1}(2-2q))$. Obviously, for finite values of $Q$, this is
only vanishing for $Q=1$, which is however not a solution of (\ref{aa}).
Therefore $Q$ does not have an extremum and property ii) holds. Property
iii) holds trivially in this case.

\subsubsection{$\mathbf{\tilde{g}\neq A}_{1}$}

For the other cases one may in principle proceed in a similar fashion, but
already for the case $\mathbf{A}_{1}\mathbf{|A}_{2}$ the analysis becomes
rather messy. For instance computing the derivative in that case, we find
that it only vanishes for $Q=(1/2\exp (mr/2(1-\exp (\sigma /2)+\sigma
/2)))^{1/(2-2q-\tilde{q})}$. Substituting this back into (\ref{aa}) we find
for a fixed value of $\sigma $ a specific value of $r$ such that the
equation is satisfied. We may then compute the second derivative and
establish that this value corresponds to a saddle point, which, in
comparison with our numerical solution exhibited in figure 1, is indeed
situated on the second plateau.

Since an analytic solution of (\ref{aa}) is eluded from our analysis so far,
we will now resort to a numerical analysis. For this purpose we discretize
the equation 
\begin{equation}
Q_{(n+1)}^{i}(r,m,\vec{\sigma})=\left( Q_{(n)}^{i}(r,m,\vec{\sigma}%
)^{2-2\exp (-m\,r/2)}+\prod_{j=1}^{\tilde{\ell}}Q_{(n)}^{j}{}(r,m,\vec{\sigma%
})^{\left[ \tilde{I}_{ij}\right] _{q_{ij}}}\right) ^{1/2}\,\,  \label{qq}
\end{equation}
and solve it iteratively in the usual fashion. Assuming convergence of this
procedure the value $n\rightarrow \infty $ is identified with the exact
solution of the recurrence relations (\ref{aa}). We start with $r=0$ and set
the initial value $Q_{0}^{j}$ to be the analytically known (see section 4)
solutions of the constant TBA-equations. Once we have achieved convergence
for a particular value of $r$, we may increase this value by an amount $%
\delta r$ and we take always as a starting value the previous solution of (%
\ref{qq}). It turns out that this procedure is extremely fast convergent
even when the particle number involved is very high. In comparison with the
full TBA equations, (\ref{qq}) are by far easier to solve since they do not
involve the complication of a convolution and correspond technically at each
value of $r$ to a constant TBA equation.

\begin{center}
\includegraphics[width=12.0cm,height=8.912cm]{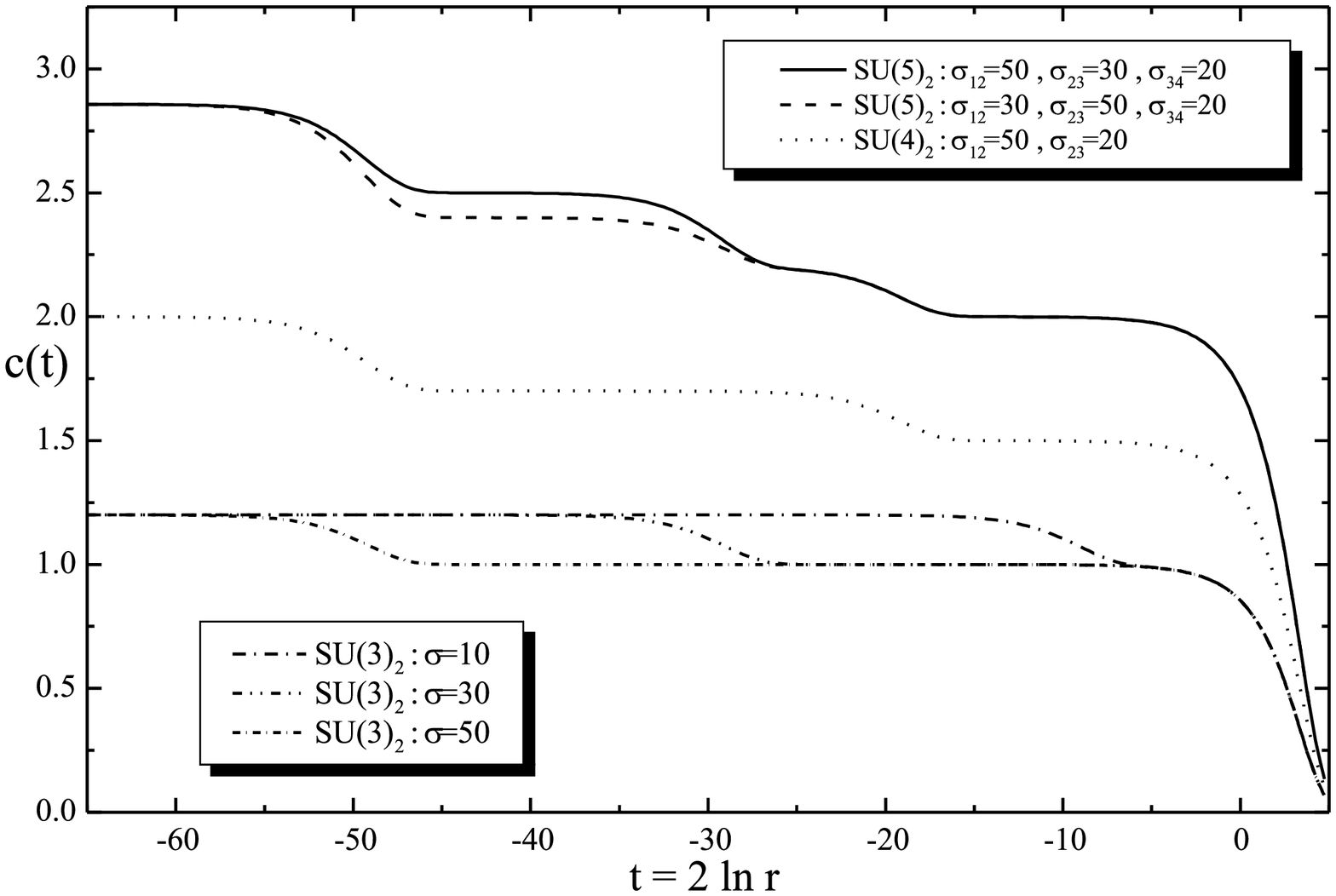}
\end{center}

\noindent {\small Figure 1: RG flow from q-deformed Virasoro characters.
\medskip }

Figure 1 shows the numerical solution of (\ref{qq}) for various algebras and
different choices of the relative order of magnitude of the resonance
parameters. We reproduce precisely the same qualitative behaviour for the
scaling function as obtained in the full TBA analysis \cite{CFKM} and from
the c-theorem \cite{CF2,CF3}. We recover all plateaux in the expected
positions. In addition we have the important property, as is seen in figure
1 for the $SU(3)_{2}$-case, that a shift in $\sigma $ by $x$ may by
compensated by a shift in $t$ with the same amount.

\subsection{$\mathbf{A}_{1}\mathbf{|E}_{6}\equiv (E_{6})_{2}$-HSG}

The approach presented in this section even allows to tackle more
complicated algebras with relatively little effort, which in the full TBA
analysis or the form factor approach constitutes a considerable
computational problem. We illustrate this by considering the $\mathbf{A}_{1}%
\mathbf{|E}_{6}$-theory. In figure 2 we present the decoupling of this
theory and report the Virasoro central charges which are taken up along the
flow as superscripts. In figure 3 we report the corresponding numerical
results of (\ref{Qr}) and (\ref{cqr}) for this theory for various different
choices of the relative order of magnitude of the resonance parameters. Our
results precisely reproduce the central charges of figure 2.

\unitlength=0.300000pt 
\begin{picture}(370.00,1000.)(-350.00,-1050.00)
\put(365.00,52.00){\makebox(0.00,0.00){{\tiny $ \alpha_6$}}}
\put(331.00,52.00){\makebox(0.00,0.00){\tiny $ \alpha_5$}}
\put(302.00,52.17){\makebox(0.00,0.00){\tiny ${\alpha}_4$}}
\put(285.00,93.00){\makebox(0.00,0.00){{\tiny ${\alpha}_2$}}}
\put(245.00,52.00){\makebox(0.00,0.00){\tiny ${\alpha}_3$}}
\put(206.00,52.00){\makebox(0.00,0.00){\tiny ${\alpha}_1$}}
\put(330.00,38.00){\line(1,0){30.00}}
\put(285.33,72.00){\line(0,-1){31.00}}
\put(290.00,38.00){\line(1,0){30.00}}
\put(250.00,38.00){\line(1,0){30.00}}
\put(210.00,38.00){\line(1,0){30.00}}
\put(285.00,78.00){\circle*{10.00}}
\put(245.00,38.00){\circle*{10.00}}
\put(325.00,38.00){\circle*{10.00}}
\put(365.00,38.00){\circle*{10.00}}
\put(285.00,38.00){\circle*{10.00}}
\put(205.00,38.00){\circle*{10.00}}
\put(258.00,-10.00){ \tiny $ \left( E_6  \right)^{\frac{36}{7}} $ }

\put(285.0,-105.00){\thicklines \line(0,1){1.00}}
\put(285.0,-85.00){\thicklines \line(0,1){10.00}}
\put(285.0,-65.00){\thicklines \line(0,1){1.00}}
\put(285.0,-45.00){\thicklines \line(0,1){10.00}}
\put(285.0,-25.00){\thicklines \line(0,1){1.00}}

\put(330.00,-162.00){\line(1,0){30.00}}
\put(290.00,-162.00){\line(1,0){30.00}}
\put(250.00,-162.00){\line(1,0){30.00}}
\put(210.00,-162.00){\line(1,0){30.00}}
\put(285.00,-122.00){\circle*{10.00}}
\put(245.00,-162.00){\circle*{10.00}}
\put(325.00,-162.00){\circle*{10.00}}
\put(365.00,-162.00){\circle*{10.00}}
\put(285.00,-162.00){\circle*{10.00}}
\put(205.00,-162.00){\circle*{10.00}}
\put(232.00,-212.00){\tiny $ \left( A_1 \otimes A_5  \right)^{\frac{17}{4}} $ }

\put(-65.0,-105.00){\thicklines \line(3,2){210.00}}

\put(-20.00,-162.00){\line(1,0){30.00}}
\put(-65.0,-128.00){\line(0,-1){31.00}}
\put(-60.00,-162.00){\line(1,0){30.00}}
\put(-100.00,-162.00){\line(1,0){30.00}}
\put(-65.00,-122.00){\circle*{10.00}}
\put(-105.00,-162.00){\circle*{10.00}}
\put(-25.00,-162.00){\circle*{10.00}}
\put(15.00,-162.00){\circle*{10.00}}
\put(-65.00,-162.00){\circle*{10.00}}
\put(-145.00,-162.00){\circle*{10.00}}
\put(-123.00,-212.00){ \tiny $ \left( A_1 \otimes D_5  \right)^{\frac{9}{2}} $ }
\thicklines
\dottedline{10}(635.0,-105.00)(425.0,35.00)

\put(680.00,-162.00){\line(1,0){30.00}}
\put(635.00,-128.00){\line(0,-1){31.00}}
\put(640.00,-162.00){\line(1,0){30.00}}
\put(560.00,-162.00){\line(1,0){30.00}}
\put(635.00,-122.00){\circle*{10.00}}
\put(595.00,-162.00){\circle*{10.00}}
\put(675.00,-162.00){\circle*{10.00}}
\put(715.00,-162.00){\circle*{10.00}}
\put(635.00,-162.00){\circle*{10.00}}
\put(555.00,-162.00){\circle*{10.00}}
\put(585.00,-212.00){\tiny  $ \left( A_2 \otimes A_4  \right)^{\frac{142}{35}} $ }

\put(285.0,-305.00){\thicklines \line(0,1){1.00}}
\put(285.0,-285.00){\thicklines \line(0,1){10.00}}
\put(285.0,-265.00){\thicklines \line(0,1){1.00}}
\put(285.0,-245.00){\thicklines \line(0,1){10.00}}
\put(285.0,-225.00){\thicklines \line(0,1){1.00}}

\put(330.00,-362.00){\line(1,0){30.00}}
\put(285.00,-328.00){\line(0,-1){31.00}}
\put(290.00,-362.00){\line(1,0){30.00}}
\put(285.00,-322.00){\circle*{10.00}}
\put(245.00,-362.00){\circle*{10.00}}
\put(325.00,-362.00){\circle*{10.00}}
\put(365.00,-362.00){\circle*{10.00}}
\put(285.00,-362.00){\circle*{10.00}}
\put(205.00,-362.00){\circle*{10.00}}
\put(215.00,-412.00){\tiny  $ \left( A_1^{\otimes 2} \otimes A_4  \right)^{\frac{27}{7}} $ }

\put(-65.0,-305.00){\thicklines \line(0,1){10.00}}
\put(-65.0,-285.00){\thicklines \line(0,1){10.00}}
\put(-65.0,-265.00){\thicklines \line(0,1){10.00}}
\put(-65.0,-245.00){\thicklines \line(0,1){10.00}}

\thicklines
\dashline{10}(-65.0,-630.00)(155.0,-725.00)

\put(10.0,-235.00){\thicklines \line(2,-1){200.00}}

\put(-65.0,-328.00){\line(0,-1){31.00}}
\put(-60.00,-362.00){\line(1,0){30.00}}
\put(-100.00,-362.00){\line(1,0){30.00}}
\put(-65.00,-322.00){\circle*{10.00}}
\put(-105.00,-362.00){\circle*{10.00}}
\put(-25.00,-362.00){\circle*{10.00}}
\put(15.00,-362.00){\circle*{10.00}}
\put(-65.00,-362.00){\circle*{10.00}}
\put(-145.00,-362.00){\circle*{10.00}}
\put(-147.00,-412.00){ \tiny $ \left( A_1^{\otimes  2} \otimes D_4  \right)^{4} $ }

\put(635.0,-305.00){\thicklines \line(0,1){1.00}}
\put(635.0,-295.00){\thicklines \line(0,1){1.00}}
\put(635.0,-285.00){\thicklines \line(0,1){1.00}}
\put(635.0,-275.00){\thicklines \line(0,1){1.00}}
\put(635.0,-265.00){\thicklines \line(0,1){1.00}}
\put(635.0,-255.00){\thicklines \line(0,1){1.00}}
\put(635.0,-245.00){\thicklines \line(0,1){1.00}}
\put(635.0,-235.00){\thicklines \line(0,1){1.00}}
\put(635.0,-225.00){\thicklines \line(0,1){1.00}}

\put(680.00,-362.00){\line(1,0){30.00}}
\put(640.00,-362.00){\line(1,0){30.00}}
\put(560.00,-362.00){\line(1,0){30.00}}
\put(635.00,-322.00){\circle*{10.00}}
\put(595.00,-362.00){\circle*{10.00}}
\put(675.00,-362.00){\circle*{10.00}}
\put(715.00,-362.00){\circle*{10.00}}
\put(635.00,-362.00){\circle*{10.00}}
\put(555.00,-362.00){\circle*{10.00}}
\put(557.00,-412.00){ \tiny $ \left( A_1 \otimes A_2 \otimes A_3   \right)^{\frac{37}{10}} $ }

\put(360.0,-425.00){\thicklines \line(2,-1){200.00}}

\put(-65.0,-505.00){\thicklines \line(0,1){10.00}}
\put(-65.0,-485.00){\thicklines \line(0,1){10.00}}
\put(-65.0,-465.00){\thicklines \line(0,1){10.00}}
\put(-65.0,-445.00){\thicklines \line(0,1){10.00}}

\put(-20.00,-562.00){\line(1,0){30.00}}
\put(-60.00,-562.00){\line(1,0){30.00}}
\put(-65.00,-522.00){\circle*{10.00}}
\put(-105.00,-562.00){\circle*{10.00}}
\put(-25.00,-562.00){\circle*{10.00}}
\put(15.00,-562.00){\circle*{10.00}}
\put(-65.00,-562.00){\circle*{10.00}}
\put(-145.00,-562.00){\circle*{10.00}}
\put(-140.00,-612.00){ \tiny $ \left( A_1^{\otimes  3} \otimes A_3  \right)^{\frac{7}{2}} $ }

\put(635.0,-505.00){\thicklines \line(0,1){1.00}}
\put(635.0,-495.00){\thicklines \line(0,1){1.00}}
\put(635.0,-485.00){\thicklines \line(0,1){1.00}}
\put(635.0,-475.00){\thicklines \line(0,1){1.00}}
\put(635.0,-465.00){\thicklines \line(0,1){1.00}}
\put(635.0,-455.00){\thicklines \line(0,1){1.00}}
\put(635.0,-445.00){\thicklines \line(0,1){1.00}}
\put(635.0,-435.00){\thicklines \line(0,1){1.00}}
\put(635.0,-425.00){\thicklines \line(0,1){1.00}}

\put(635.0,-630.00){\thicklines \line(-2,-1){210.00}}

\put(680.00,-562.00){\line(1,0){30.00}}
\put(560.00,-562.00){\line(1,0){30.00}}
\put(635.00,-522.00){\circle*{10.00}}
\put(595.00,-562.00){\circle*{10.00}}
\put(675.00,-562.00){\circle*{10.00}}
\put(715.00,-562.00){\circle*{10.00}}
\put(635.00,-562.00){\circle*{10.00}}
\put(555.00,-562.00){\circle*{10.00}}
\put(567.00,-612.00){\tiny  $ \left( A_1^{\otimes  2} \otimes A_2^{\otimes  2}   
   \right)^{\frac{17}{5}} $ }

\put(210.00,-762.00){\line(1,0){30.00}}
\put(285.00,-722.00){\circle*{10.00}}
\put(245.00,-762.00){\circle*{10.00}}
\put(325.00,-762.00){\circle*{10.00}}
\put(365.00,-762.00){\circle*{10.00}}
\put(285.00,-762.00){\circle*{10.00}}
\put(205.00,-762.00){\circle*{10.00}}
\put(210.00,-812.00){ \tiny $ \left( A_1^{\otimes  4} \otimes A_2 
   \right)^{\frac{16}{5}} $ }

\put(285.0,-905.00){\thicklines \line(0,1){60.00}}

\put(285.00,-922.00){\circle*{10.00}}
\put(245.00,-962.00){\circle*{10.00}}
\put(325.00,-962.00){\circle*{10.00}}
\put(365.00,-962.00){\circle*{10.00}}
\put(285.00,-962.00){\circle*{10.00}}
\put(205.00,-962.00){\circle*{10.00}}
\put(245.00,-1012.00){\tiny  $ \left( A_1^{ \otimes 6}  \right)^{3} $ }
\end{picture}

\noindent {\small Figure 2: The decoupling of the }$A_{1}|E_{6}${\small %
-theory.}

\begin{center}
\includegraphics[width=12.0cm,height=8.912cm]{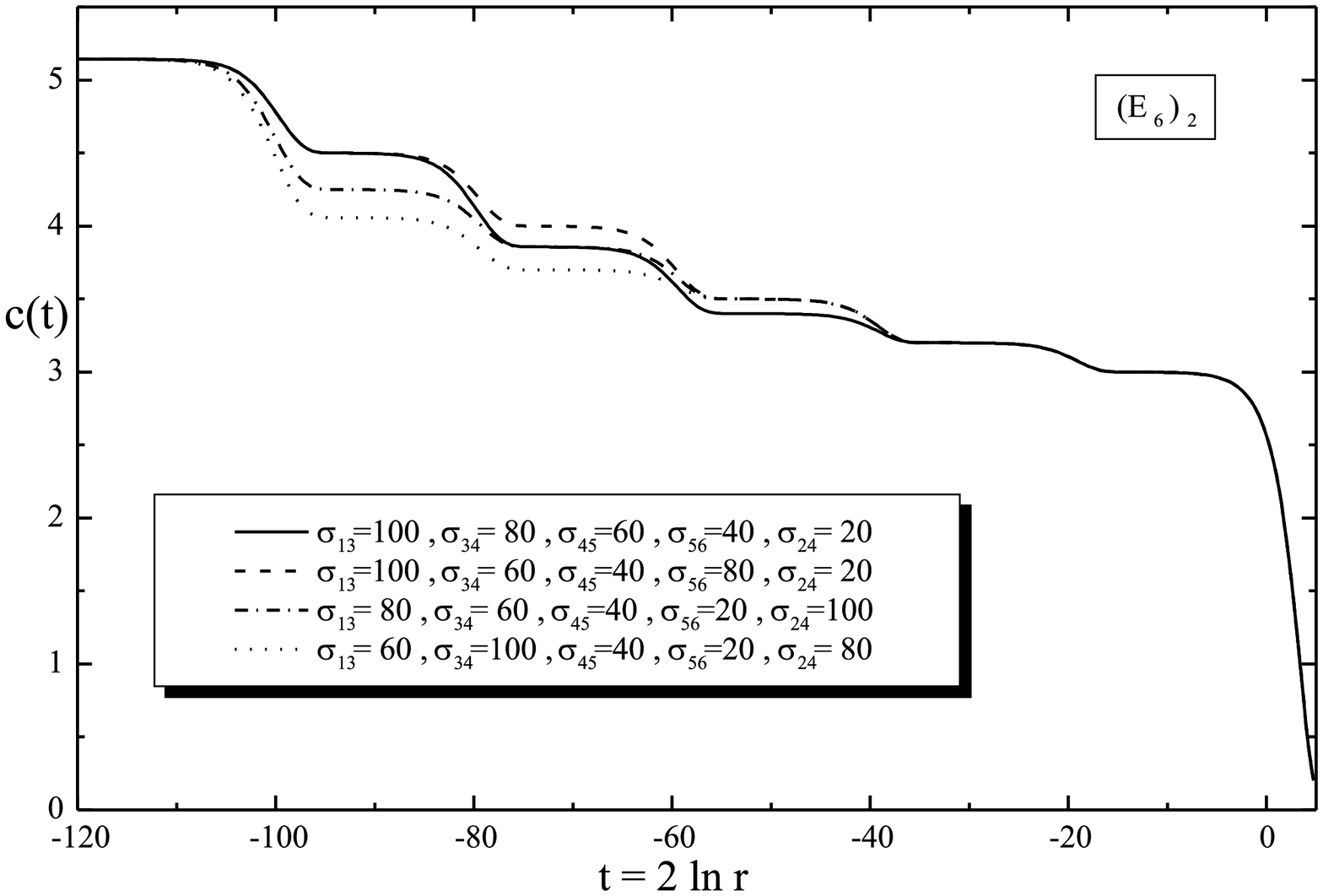}
\end{center}

\noindent {\small Figure 3: RG flow from q-deformed Virasoro characters.}

\subsection{\textbf{g\TEXTsymbol{\vert}\~{g}}--coset theories}

Recalling now from section 2.2 the various ways in which we can represent
the unitary series, we may construct the flows between different cosets in a
similar way as in the preceding subsection for a single homogeneous
sine-Gordon theory. Figure 4 exhibits the flow along the unitary series of
the $N=0,1,2$ superconformal minimal models.

From the realizations of the various cosets in terms of HSG-models it is
also clear that we may produce flows between the different series as
suggested in \cite{Foda} by alternative means. By controlling the energy
scale of the unstable particle we obtain 
\begin{eqnarray*}
\mathcal{M}^{N=2}(k,k+1) &\equiv &A_{1}|D_{k+2}/A_{1}|A_{k-1}\otimes
A_{1}|A_{1}^{\otimes 2}\quad \stackunder{\sigma _{k+1,k+2}\rightarrow \infty 
}{\longrightarrow } \\
\mathcal{M}^{N=1}(k,k+1) &\equiv &A_{1}|A_{k+1}/A_{1}|A_{k-1}\otimes
A_{1}|A_{1}\quad \quad \stackunder{\sigma _{k,k+1}\rightarrow \infty }{%
\longrightarrow } \\
\mathcal{M}(k,k+1) &\equiv &A_{1}|A_{k}/A_{1}|A_{k-1}\quad .
\end{eqnarray*}
Our numerical results which reproduce these flows are presented in figure 5.
It is this type of flow which in\ \cite{Foda} was realized as so-called
``Bailey flow''.

\bigskip \bigskip\bigskip

\begin{center}
\includegraphics[width=12.0cm,height=8.912cm]{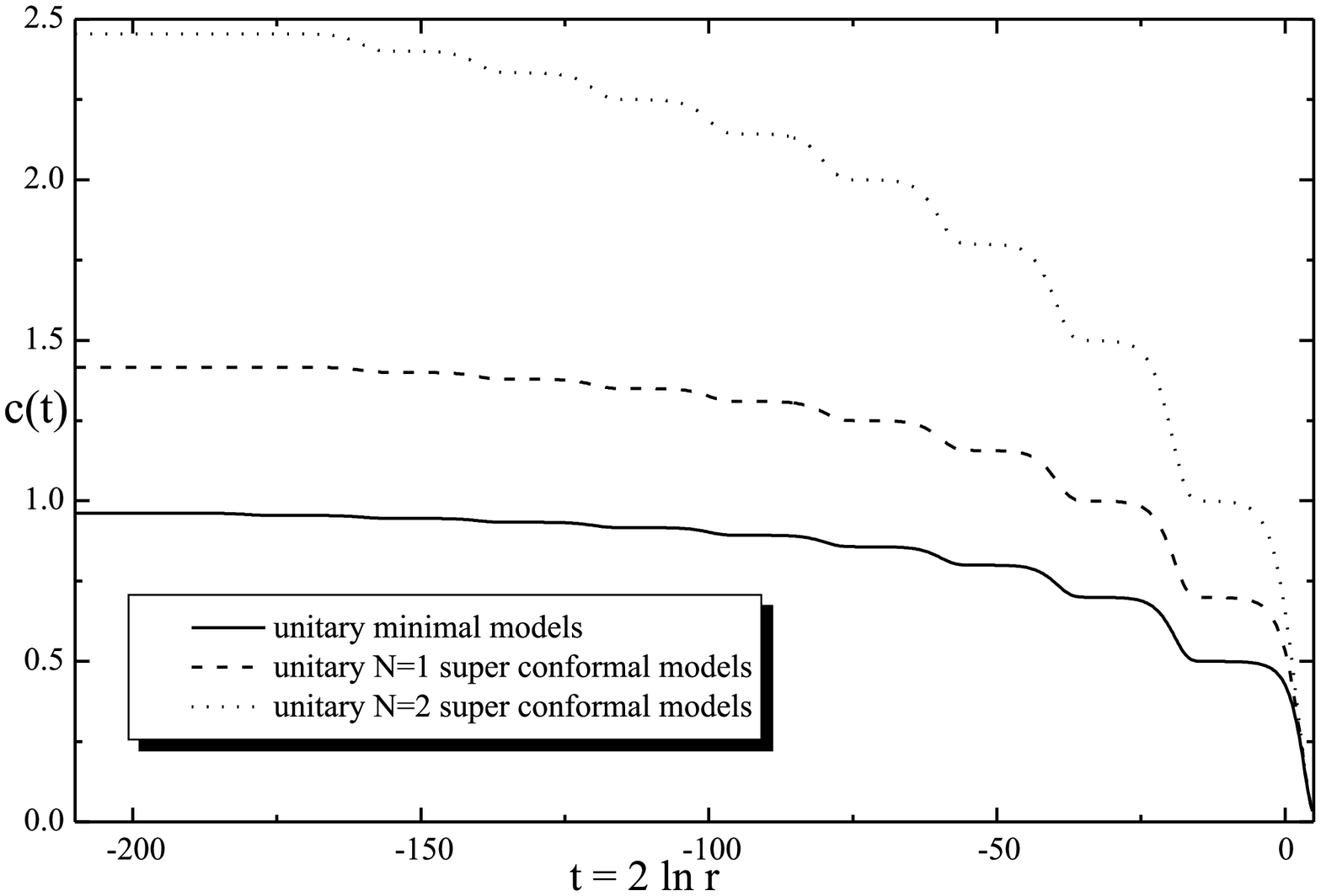}
\end{center}

\noindent {\small Figure 4: Internal RG flow for the }$N=0,1,2${\small \
unitary minimal models.}

\begin{center}
\includegraphics[width=12.0cm,height=8.912cm]{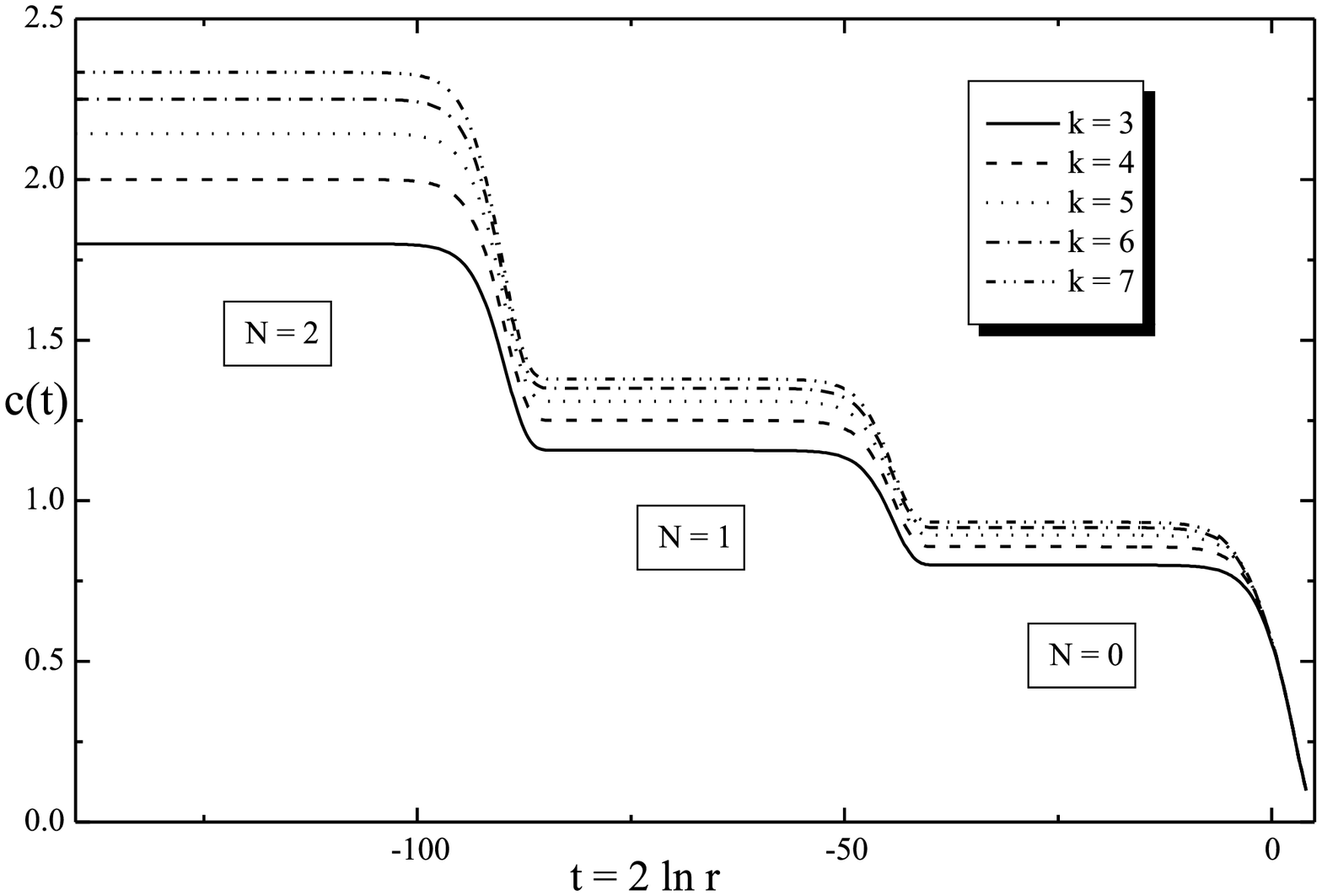}
\end{center}

\noindent {\small Figure 5: RG flow between }$N=0,1,2${\small \ unitary
minimal models.}

\section{The fixed point solutions}

As we have mentioned, we do not have a general solution of (\ref{Qr}) so far
for the entire range of $r$, but at each fixed point such expressions may be
found. In \cite{Resh,Kun} it was noted, that the recurrence relations (\ref
{cTBA}) admit closed analytical solutions in terms of some very distinct
mathematical objects, namely Weyl characters. Since the proofs of these
identities are very often missing or only indicated in the literature, we
find it instructive to present various transparent proofs in this section.
In addition we present numerous new solutions for theories treated before
and for some hitherto not considered at all. We start by assembling several
properties of the characters which we utilize later to solve the recurrence
relations (\ref{cTBA}) or equivalently (\ref{Qr}) in the range of $r$
characterized by property iii) in the introduction.

\subsection{Properties of Weyl Characters}

The characters for the representation of a simple Lie algebra \textbf{g }%
with rank $\ell $ are well-known to be expressible in terms of the famous
Weyl character formula, see e.g. \cite{Hum}. From all the equivalent
formulations of this formula the version 
\begin{equation}
\chi _{\lambda }(q)=\prod\limits_{\alpha \in \Delta _{+}}\frac{q^{\alpha
\cdot (\lambda +\rho )}-q^{-\alpha \cdot (\lambda +\rho )}}{q^{\alpha \cdot
\rho }-q^{-\alpha \cdot \rho }}\,\,  \label{Weyl}
\end{equation}
is most convenient for our purposes. Here $\lambda $ denotes an arbitrary
weight, $\Delta _{+}$ the set of positive roots and $\rho =1/2\sum_{\alpha
\in \Delta _{+}}\alpha =\sum_{i=1}^{\ell }\lambda _{i}$ the Weyl vector with 
$\lambda _{i}$ denoting the fundamental weights. Needless to say that like (%
\ref{chi}) the $\chi _{\lambda }(q)$'s constitute generating \ functions in
the formal parameter $q$ which is unrelated to the deformation parameter of
section 3. We will be particularly interested in the form of (\ref{Weyl})
evaluated at the special value $q=e^{i\pi \tau }$ 
\begin{equation}
\chi _{\lambda }(\tau )=\dprod\limits_{\alpha \in \Delta _{+}}\frac{\sin
\left( \alpha \cdot (\lambda +\rho )\pi \tau \right) }{\sin \left( \alpha
\cdot \rho \pi \tau \right) }\,\,.  \label{ww}
\end{equation}
When considering $\lambda $ to be a fundamental weight $\lambda _{i}$, it is
useful to employ the conventions $\chi _{\lambda _{0}}=\chi _{\lambda _{\ell
+1}}=1$ and set $\chi _{\lambda _{-n}}=0$ for a positive integer $n$. When $%
\tau $ approaches 0, we obtain the well-known formula for the dimension of
the particular representation of the weight $\lambda $%
\begin{equation}
\dim \lambda =\prod\limits_{\alpha \in \Delta _{+}}\frac{\alpha \cdot
(\lambda +\rho )}{\alpha \cdot \rho }\,\,.
\end{equation}

We now wish to establish various properties for the character $\chi
_{\lambda }(\tau )$. It appears difficult to carry out these studies on the
generic expression (\ref{ww}) and we shall therefore resort to a
case-by-case analysis. Denoting by $\varepsilon _{1},\ldots ,\varepsilon
_{n} $ the standard orthonormal basis of $\Bbb{R}^{n}$ with $\varepsilon
_{i}\cdot \varepsilon _{j}=\delta _{ij}$, it is well-known that it is
possible to represent the entire root system as vectors on a suitably chosen
lattice in $\Bbb{R}^{n}$ with one (simply laced) or two (non-simply laced)
prescribed lengths. We adopt the conventions of Bourbaki \cite{Bour}, which
resulted historically from an investigation of the adjoint representation of
simple Lie algebras, which is the reason why they appear not always entirely
obvious.

\subsubsection{\textbf{A}$_{\ell }$}

\unitlength=0.680000pt 
\begin{picture}(437.92,75.00)(50.00,95.00)
\qbezier(260.00,140.00)(365.00,95.00)(471.00,140.00)
\qbezier(301.00,140.00)(365.00,110.00)(431.00,140.00)
\qbezier(341.00,140.00)(365.00,125.00)(391.00,140.00)
\put(396.00,165.00){\makebox(0.00,0.00){${\alpha}_{\ell-2}$}}
\put(436.00,165.00){\makebox(0.00,0.00){${\alpha}_{\ell-1}$}}
\put(476.00,165.00){\makebox(0.00,0.00){${\alpha}_{\ell}$}}
\put(336.00,165.00){\makebox(0.00,0.00){${\alpha}_3$}}
\put(296.00,165.00){\makebox(0.00,0.00){${\alpha}_2$}}
\put(255.00,165.00){\makebox(0.00,0.00){${\alpha}_1$}}
\put(400.00,150.00){\line(1,0){30.00}}
\put(440.00,150.00){\line(1,0){30.00}}
\put(390.00,150.00){\line(-1,0){10.00}}
\put(340.00,150.00){\line(1,0){10.00}}
\put(300.00,150.00){\line(1,0){30.00}}
\put(260.00,150.00){\line(1,0){30.00}}
\put(395.00,150.00){\circle*{10.00}}
\put(435.00,150.00){\circle*{10.00}}
\put(475.00,150.00){\circle*{10.00}}
\put(255.00,150.00){\circle*{10.00}}
\put(295.00,150.00){\circle*{10.00}}
\put(335.00,150.00){\circle*{10.00}}
\end{picture}

\noindent We represent the roots of $A_{\ell }$ in $\Bbb{R}^{\ell +1}$.
According to \cite{Bour} all positive roots are given by 
\begin{equation}
\varepsilon _{i}-\varepsilon _{j}=\alpha \in \Delta _{+}\qquad \qquad \quad
\quad \text{for }1\leq i<j\leq \ell +1\,\,\,.
\end{equation}
The fundamental weights and the Weyl vector are realized as 
\begin{equation}
\lambda _{k}=\sum\limits_{i=1}^{k}\varepsilon _{i}-\frac{k}{\ell +1}%
\sum\limits_{i=1}^{\ell +1}\varepsilon _{i}\quad \text{and \quad }\rho
=\sum\limits_{i=1}^{\ell +1}(\ell /2+1-i)\,\varepsilon _{i}\,\,.
\end{equation}
Equipped with these quantities we can evaluate (\ref{ww}) and obtain more
explicit formulae 
\begin{equation}
\chi _{a\lambda _{k}}(\tau )=\prod\limits_{1\leq i<j\leq \ell +1}\frac{\sin
[(\varepsilon _{i}-\varepsilon _{j})\cdot (a\lambda _{k}+\rho )\pi \tau ]}{%
\sin [(\varepsilon _{i}-\varepsilon _{j})\cdot \rho \pi \tau ]}%
=\prod\limits_{i=1}^{k}\prod\limits_{j=k}^{\ell }\frac{\sin [(a+1+j-i)\pi
\tau ]}{\sin [(1+j-i)\pi \tau ]}\,\,.  \label{exAA}
\end{equation}
The last expression in (\ref{exAA}) is best suited to establish various
properties of the $A_{\ell }$ related characters 
\begin{eqnarray}
\chi _{a\lambda _{k}}(\tau ) &=&\chi _{a\lambda _{k}}(\tau +2)  \label{Aid}
\\
\chi _{a\lambda _{k}}(\tau ) &=&\chi _{a\lambda _{\ell +1-k}}(\tau )
\label{Aid2} \\
\chi _{(a+1)\lambda _{k}}(\tau ) &=&\chi _{a\lambda _{k}}(\tau
)\prod\limits_{j=1}^{k}\frac{\sin [(a+\ell +2-j)\pi \tau ]}{\sin
[(a+k+1-j)\pi \tau ]}  \label{h1} \\
\chi _{a\lambda _{k+1}}(\tau ) &=&\chi _{a\lambda _{k}}(\tau
)\prod\limits_{j=1+k}^{\ell }\frac{\sin [(a+j)\pi \tau ]}{\sin [j\pi \tau ]}%
\prod\limits_{i=1}^{k}\frac{\sin [(\ell +1-j)\pi \tau ]}{\sin [(a+\ell
+1-j)\pi \tau ]}  \label{h2} \\
\chi _{a\lambda _{k}}(\tau )\chi _{a\lambda _{k}}(\tau ) &=&\chi
_{(a+1)\lambda _{k}}(\tau )\chi _{(a-1)\lambda _{k}}(\tau )+\chi _{a\lambda
_{k+1}}(\tau )\chi _{a\lambda _{k-1}}(\tau )\,\,.  \label{h3}
\end{eqnarray}
Here (\ref{Aid}) is obvious and (\ref{Aid2}), (\ref{h1}), (\ref{h2}) follow
from simple shifts in (\ref{exAA}). With the help of (\ref{h1}) and (\ref{h2}%
) we can verify (\ref{h3}). Note that (\ref{Aid})-(\ref{h3}) hold for
generic values of $\tau $. We now also want to identify $\chi _{a\lambda
_{k}}$ and $\chi _{(\tilde{l}+1-a)\lambda _{k}}$ for some integer $\tilde{l}$%
. This is, however, not true for generic values of $\tau $. Expressing $\chi
_{a\lambda _{k}}$ and $\chi _{(\tilde{l}+1-a)\lambda _{k}}$ in the form (\ref
{exAA}) and denoting the variables over which the products are taken in the
former by $i,j$ and the latter by $i^{\prime },j^{\prime }$ the two
characters obviously coincide if $(a+j-i)\tau =1+(a-\tilde{l}-1-j^{\prime
}+i^{\prime })\tau $. From the available values of $i,j,i^{\prime
},j^{\prime }$ the combination $j+j^{\prime }-i-i^{\prime }=\ell +1$
constitutes a consistent solution of this equation such that we have 
\begin{equation}
\chi _{a\lambda _{k}}(\tau =\tfrac{1}{\ell +\tilde{\ell}+2})=\chi _{(\tilde{l%
}+1-a)\lambda _{k}}(\tau =\tfrac{1}{\ell +\tilde{\ell}+2})\,\,.
\end{equation}
This means it is the symmetry of the Dynkin diagram which fixes the value of 
$\tau $.

\subsubsection{\textbf{D}$_{\ell }$}

\unitlength=0.680000pt 
\begin{picture}(437.92,130.00)(50.00,65.00)
\qbezier(420.00,170.00)(440.00,150.00)(420.00,130.00)
\put(437.50,104.17){\makebox(0.00,0.00){$\alpha_{\ell}$}}
\put(437.92,184.17){\makebox(0.00,0.00){${\alpha}_{\ell -1}$}}
\put(411.17,149.59){\makebox(0.00,0.00){${\alpha}_{\ell-2}$}}
\put(347.50,165.00){\makebox(0.00,0.00){${\alpha}_{\ell-3}$}}
\put(296.00,165.00){\makebox(0.00,0.00){${\alpha}_2$}}
\put(255.00,165.00){\makebox(0.00,0.00){${\alpha}_1$}}
\put(389.00,146.33){\line(1,-1){22.67}}
\put(411.67,176.67){\line(-1,-1){23.00}}
\put(350.00,150.00){\line(1,0){30.00}}
\put(330.00,150.00){\line(1,0){10.00}}
\put(300.00,150.00){\line(1,0){10.00}}
\put(260.00,150.00){\line(1,0){30.00}}
\put(415.00,120.00){\circle*{10.00}}
\put(415.00,180.00){\circle*{10.00}}
\put(385.00,150.00){\circle*{10.00}}
\put(345.00,150.00){\circle*{10.00}}
\put(255.00,150.00){\circle*{10.00}}
\put(295.00,150.00){\circle*{10.00}}
\end{picture}

\noindent We represent the roots of $D_{\ell }$ in $\Bbb{R}^{\ell }$.
According to \cite{Bour} all positive roots are expressible as 
\begin{equation}
\varepsilon _{i}\pm \varepsilon _{j}=\alpha \in \Delta _{+}\qquad \qquad 
\text{for }1\leq i<j\leq \ell \quad .
\end{equation}
The fundamental weights are given by 
\begin{equation}
\lambda _{\ell -1}=\sum\limits_{i=1}^{\ell -1}\frac{\varepsilon
_{i}-\varepsilon _{\ell }}{2},\qquad \lambda _{\ell }=\frac{1}{2}%
\sum\limits_{i=1}^{\ell }\,\varepsilon _{i}\,\,,\quad \lambda
_{k}=\sum\limits_{i=1}^{k}\,\varepsilon _{i}\,\,\quad \text{for }1\leq k\leq
\ell -2,
\end{equation}
such that the Weyl vector reads 
\begin{equation}
\rho =\sum\limits_{i=1}^{\ell -1}\,(\ell -i)\varepsilon _{i}\,\,.
\end{equation}
Substituting these quantities into (\ref{ww}) yields 
\begin{equation}
\chi _{a\lambda _{k}}(\tau )=\prod\limits_{1\leq i<j\leq \ell }\frac{\sin
[(\varepsilon _{i}-\varepsilon _{j})\cdot (a\lambda _{k}+\rho )\pi \tau ]}{%
\sin [(\varepsilon _{i}-\varepsilon _{j})\cdot \rho \pi \tau ]}\frac{\sin
[(\varepsilon _{i}+\varepsilon _{j})\cdot (a\lambda _{k}+\rho )\pi \tau ]}{%
\sin [(\varepsilon _{i}+\varepsilon _{j})\cdot \rho \pi \tau ]}
\end{equation}
from which we derive 
\begin{eqnarray}
\chi _{a\lambda _{k}}(\tau )\!\! &=&\!\!\!\!\!\!\!\prod\limits_{1\leq
i<j\leq k}\!\!\!\!\tfrac{\sin [(2a+2\ell -i-j)\pi \tau ]}{\sin [(2\ell
-i-j)\pi \tau ]}\prod\limits_{i=1}^{k}\prod\limits_{j=k+1}^{\ell }\!\!\tfrac{%
\sin [(a+j-i)\pi \tau ]}{\sin [(j-i)\pi \tau ]}\tfrac{\sin [(2\ell
+a-j-i)\pi \tau ]}{\sin [(2\ell -j-i)\pi \tau ]},\text{ }{\small 1\leq k\leq
\ell -2}\,\,\,\,\,\,\,\,\,\,\,  \label{11} \\
\chi _{a\lambda _{\ell }}(\tau ) &=&\chi _{a\lambda _{\ell -1}}(\tau
)=\prod\limits_{1\leq i<j\leq \ell }\tfrac{\sin [(2\ell +a-i-j)\pi \tau ]}{%
\sin [(2\ell -i-j)\pi \tau ]}\,\,.  \label{22}
\end{eqnarray}
From (\ref{11}) and (\ref{22}) we can now deduce various properties of the $%
D_{\ell }$ related characters 
\begin{eqnarray}
\chi _{a\lambda _{k}}(\tau ) &=&\chi _{a\lambda _{k}}(\tau +2), \\
\chi _{a\lambda _{\ell }}(\tau ) &=&\chi _{a\lambda _{\ell -1}}(\tau ), \\
\chi _{a\lambda _{1}}(\tau ^{\prime }) &=&\sum_{k=0}^{\infty }(-1)^{k}\chi
_{\lambda _{a-2k}}(\tau ^{\prime }),\,\quad \quad a\leq \ell -2, \\
\chi _{\lambda _{n+1}}(\tau ^{\prime }) &=&\chi _{(n+1)\lambda _{1}}(\tau
^{\prime })+\chi _{(n-1)\lambda _{1}}(\tau ^{\prime }), \\
\chi _{\lambda _{1}}(\tau ^{\prime })\chi _{\lambda _{1}}(\tau ^{\prime })
&=&2\chi _{\lambda _{2}}(\tau ^{\prime }), \\
\chi _{\lambda _{\ell }}(\tau ^{\prime })\chi _{\lambda _{\ell }}(\tau
^{\prime }) &=&2\sum_{k=0}^{\infty }\chi _{\lambda _{\ell -2-4k}}(\tau
^{\prime }).
\end{eqnarray}
\noindent Here we have set $\tau ^{\prime }=1/(4\ell -4)$.

\subsubsection{\textbf{E}$_{6}$}

\unitlength=0.680000pt 
\begin{picture}(370.00,100.0)(0.00,-20.00)
\qbezier(211.00,28.00)(285.00,-7.00)(360.00,28.00)
\qbezier(251.00,28.00)(285.00,8.0)(320.00,28.00)
\put(365.00,52.00){\makebox(0.00,0.00){$\alpha_6$}}
\put(325.00,52.00){\makebox(0.00,0.00){$\alpha_5$}}
\put(296.25,52.17){\makebox(0.00,0.00){${\alpha}_4$}}
\put(285.00,93.00){\makebox(0.00,0.00){${\alpha}_2$}}
\put(245.00,52.00){\makebox(0.00,0.00){${\alpha}_3$}}
\put(206.00,52.00){\makebox(0.00,0.00){${\alpha}_1$}}
\put(330.00,38.00){\line(1,0){30.00}}
\put(285.33,72.67){\line(0,-1){31.00}}
\put(290.00,38.00){\line(1,0){30.00}}
\put(250.00,38.00){\line(1,0){30.00}}
\put(210.00,38.00){\line(1,0){30.00}}
\put(285.00,78.00){\circle*{10.00}}
\put(245.00,38.00){\circle*{10.00}}
\put(325.00,38.00){\circle*{10.00}}
\put(365.00,38.00){\circle*{10.00}}
\put(285.00,38.00){\circle*{10.00}}
\put(205.00,38.00){\circle*{10.00}}
\end{picture}

\noindent Following still \cite{Bour} the roots and weights of $E_{6}$ may
be represented in $\Bbb{R}^{8}$, where we label the roots as depicted in the
preceding Dynkin diagram. Since these expressions are rather cumbersome, we
refer the reader to the literature and report here only the final
expressions for the characters. Noting that all characters are of the
general form $\prod_{1\leq x<h}\sin (\pi \tau (a+x))/\sin (\pi \tau x)$,
with $h$ being the Coxeter number, it is convenient to use the following
notation 
\begin{equation}
\{a_{1,1}^{x_{1,1}},\ldots a_{1,b_{1}}^{x_{1,b_{1}}};\ldots
;a_{i,1}^{x_{i,1}},a_{i,2}^{x_{i,2}}\ldots ,a_{i,b_{i}}^{x_{i,b_{i}}},\ldots
\}:=\prod_{i=1}^{h-1}\prod_{j=1}^{b_{i}}\left( \frac{\sin \pi \tau
(a_{i,j}+i)}{\sin \pi \tau i}\right) ^{x_{i,j}}\,.
\end{equation}
Note that all expressions we find have at least one $x_{i,j}\neq 0$ for each 
$i\in \lbrack 1,h-1]$. We compute 
\begin{eqnarray}
&&\chi _{a\lambda _{1}}=\{a;a;a;a^{2};a^{2};a^{2};a^{2};a^{2};a;a;a\}
\label{E1} \\
&&\chi _{a\lambda _{2}}=\{a;a;a^{2};a^{3};a^{3};a^{3};a^{3};a^{2};a;a;2a\} \\
&&\chi _{a\lambda
_{3}}=\{a;a^{2};a^{3};a^{4};a^{4};a^{3};a^{2},2a;a,2a;2a;2a;2a\} \\
&&\chi _{a\lambda
_{4}}=\{a;a^{3};a^{5};a^{5};a^{3},2a;a,2a^{2};2a^{3};2a^{2};2a;3a;3a\}\,\,.%
\quad  \label{E4}
\end{eqnarray}
Here and in the following we suppress the $\tau $-dependence of $\chi $,
i.e. we read $\chi _{a\lambda i}=\chi _{a\lambda _{i}}(\tau )$.

\subsubsection{\textbf{E}$_{7}$}

\unitlength=0.680000pt 
\begin{picture}(370.00,107.58)(0.00,0.00)
\put(405.00,52.00){\makebox(0.00,0.00){$\alpha_7$}}
\put(365.00,52.00){\makebox(0.00,0.00){$\alpha_6$}}
\put(325.00,52.00){\makebox(0.00,0.00){$\alpha_5$}}
\put(296.25,52.17){\makebox(0.00,0.00){${\alpha}_4$}}
\put(285.00,93.00){\makebox(0.00,0.00){${\alpha}_2$}}
\put(245.00,52.00){\makebox(0.00,0.00){${\alpha}_3$}}
\put(206.00,52.00){\makebox(0.00,0.00){${\alpha}_1$}}
\put(370.00,38.00){\line(1,0){30.00}}
\put(330.00,38.00){\line(1,0){30.00}}
\put(285.33,72.67){\line(0,-1){31.00}}
\put(290.00,38.00){\line(1,0){30.00}}
\put(250.00,38.00){\line(1,0){30.00}}
\put(210.00,38.00){\line(1,0){30.00}}
\put(285.00,78.00){\circle*{10.00}}
\put(245.00,38.00){\circle*{10.00}}
\put(325.00,38.00){\circle*{10.00}}
\put(365.00,38.00){\circle*{10.00}}
\put(405.00,38.00){\circle*{10.00}}
\put(285.00,38.00){\circle*{10.00}}
\put(205.00,38.00){\circle*{10.00}}
\end{picture}

\noindent Our convention for naming the roots are the same as in \cite{Bour}
according to which we represent the roots of $E_{7}$ in $\Bbb{R}^{8}$. We
then compute 
\begin{eqnarray}
\chi _{a\lambda _{1}}
&=&%
\{a;a;a;a^{2};a^{2};a^{3};a^{3};a^{3};a^{3};a^{3};a^{3};a^{2};a^{2};a;a;a;2a%
\}  \label{E11} \\
\chi _{a\lambda _{2}}
&=&%
\{a;a;a^{2};a^{3};a^{4};a^{4};a^{5};a^{4};a^{4};a^{3};a^{2},2a;a,2a;a,2a;2a;2a;2a;2a\}
\\
\chi _{a\lambda _{3}}
&=&%
\{a;a^{2};a^{3};a^{4};a^{5};a^{5};a^{4},2a;a^{3},2a;a^{2},2a^{2};a,2a^{2};2a^{3};2a^{2};2a^{2};2a;2a;3a;3a\}\quad \quad
\\
\chi _{a\lambda _{4}}
&=&%
\{a;a^{3};a^{5};a^{6};a^{5},2a;a^{3},2a^{2};a,2a^{4};2a^{4};2a^{4};2a^{2},3a;2a,3a^{2};3a^{2};3a^{2};3a;4a;
\nonumber \\
&&4a;4a\} \\
\chi _{a\lambda _{5}}
&=&%
\{a;a^{2};a^{4};a^{5};a^{6};a^{5};a^{4},2a;a^{2},2a^{2};a,2a^{3};2a^{3};2a^{3};2a^{2};2a,3a;3a;3a;3a;3a\}
\\
\chi _{a\lambda _{6}}
&=&%
\{a;a^{2};a^{2};a^{3};a^{4};a^{4};a^{4};a^{4};a^{3},2a;a^{2},2a;a^{2},2a;a,2a;2a^{2};2a;2a;2a;2a\}
\\
\chi _{a\lambda _{7}}
&=&\{a;a;a;a;a^{2};a^{2};a^{2};a^{2};a^{3};a^{2};a^{2};a^{2};a^{2};a;a;a;a\}
\label{E77}
\end{eqnarray}

\noindent

\subsubsection{\textbf{E}$_{8}$}

\unitlength=0.680000pt 
\begin{picture}(370.00,107.58)(0.00,0.00)
\put(445.00,52.00){\makebox(0.00,0.00){$\alpha_8$}}
\put(405.00,52.00){\makebox(0.00,0.00){$\alpha_7$}}
\put(365.00,52.00){\makebox(0.00,0.00){$\alpha_6$}}
\put(325.00,52.00){\makebox(0.00,0.00){$\alpha_5$}}
\put(296.25,52.17){\makebox(0.00,0.00){${\alpha}_4$}}
\put(285.00,93.00){\makebox(0.00,0.00){${\alpha}_2$}}
\put(245.00,52.00){\makebox(0.00,0.00){${\alpha}_3$}}
\put(206.00,52.00){\makebox(0.00,0.00){${\alpha}_1$}}
\put(410.00,38.00){\line(1,0){30.00}}
\put(370.00,38.00){\line(1,0){30.00}}
\put(330.00,38.00){\line(1,0){30.00}}
\put(285.33,72.67){\line(0,-1){31.00}}
\put(290.00,38.00){\line(1,0){30.00}}
\put(250.00,38.00){\line(1,0){30.00}}
\put(210.00,38.00){\line(1,0){30.00}}
\put(285.00,78.00){\circle*{10.00}}
\put(245.00,38.00){\circle*{10.00}}
\put(325.00,38.00){\circle*{10.00}}
\put(365.00,38.00){\circle*{10.00}}
\put(405.00,38.00){\circle*{10.00}}
\put(445.00,38.00){\circle*{10.00}}
\put(285.00,38.00){\circle*{10.00}}
\put(205.00,38.00){\circle*{10.00}}
\end{picture}

\noindent Our convention for naming the roots are as in \cite{Bour}
according to which we represent the roots of $E_{8}$ in $\Bbb{R}^{8}$. We
compute

\begin{eqnarray}
\chi _{a\lambda _{1}}
&=&%
\{a;a;a;a^{2};a^{2};a^{3};a^{4};a^{4};a^{4};a^{5};a^{5};a^{5};a^{5};a^{4};a^{4};a^{4};a^{3},2a;a^{2},2a;a^{2},2a;
\nonumber \\
&&a,2a;a,2a;a,2a;2a^{2};2a;2a;2a;2a;2a;2a\}  \label{EE81} \\
\chi _{a\lambda _{2}}
&=&%
\{a;a;a^{2};a^{3};a^{4};a^{5};a^{6};a^{6};a^{6};a^{6};a^{5},2a;a^{4},2a;a^{3},2a^{2};a^{2},2a^{2};a,2a^{3};
\nonumber \\
&&a,2a^{3};2a^{4};2a^{3};2a^{3};2a^{2};2a^{2},2a;3a;2a;3a;3a;3a;3a;3a;3a;3a\}
\\
\chi _{a\lambda _{3}}
&=&%
\{a;a^{2};a^{3};a^{4};a^{5};a^{6};a^{6},2a;a^{5},2a;a^{4},2a^{2};a^{3},2a^{3};a^{2},2a^{4};a,2a^{4};2a^{5};2a^{4};
\nonumber \\
&&2a^{4};2a^{3},3a;2a^{2},3a;2a,3a^{2};2a,3a^{2};3a^{2};3a^{2};3a^{2};3a,4a;4a;4a;4a;4a;
\nonumber \\
&&4a;4a\}\,\, \\
\chi _{a\lambda _{4}}
&=&%
\{a;a^{3};a^{5};a^{6};a^{6},2a;a^{5},2a^{2};a^{3},2a^{4};a,2a^{5};2a^{6};2a^{5},3a;2a^{4},3a^{2};2a^{2},
\nonumber \\
&&3a^{3};2a,3a^{4};3a^{4};3a^{3},4a;3a^{2},4a^{2};3a,4a^{3};4a^{3};4a^{3};4a^{2};4a,5a;5a^{2};5a^{2};
\nonumber \\
&&5a;6a;6a;6a;6a;6a\} \\
\chi _{a\lambda _{5}}
&=&%
\{a;a^{2};a^{4};a^{6};a^{7};a^{7};a^{6},2a;a^{4},2a^{2};a^{2},2a^{4};a,2a^{5};2a^{6};2a^{5};2a^{4},3a;2a^{2},
\nonumber \\
&&3a^{2};2a^{2},3a^{3};3a^{4};3a^{4};3a^{3};3a^{2},4a;3a,4a;4a^{2};4a^{2};4a^{2};4a;5a;5a;5a;5a\}
\\
\chi _{a\lambda _{6}}
&=&%
\{a;a^{2};a^{3};a^{4};a^{5};a^{6};a^{6};a^{6};a^{5},2a;a^{4},2a^{2};a^{3},2a^{3};a^{2},2a^{3};a,2a^{4};2a^{4};
\nonumber \\
&&2a^{4};2a^{3},3a;2a^{3},3a;2a^{2},3a;2a,3a^{2};3a^{2};3a^{2};3a^{2};3a^{2};3a;3a;3a;4a;
\nonumber \\
&&4a,;4a\} \\
\chi _{a\lambda _{7}}
&=&%
\{a;a^{2};a^{2};a^{2};a^{3};a^{4};a^{4};a^{4};a^{5};a^{5};a^{4},2a;a^{4},2a;a^{4},2a;a^{3},2a;a^{2},2a^{2};
\nonumber \\
&&a^{2},2a^{2};a^{2},2a^{2};a,2a^{2};2a^{3};2a^{2};2a^{2};2a^{2};2a^{2};2a;2a;2a;2a;3a;3a\}
\\
\chi _{a\lambda _{8}}
&=&%
\{a;a;a;a;a;a^{2};a^{2};a^{2};a^{2};a^{3};a^{3};a^{3};a^{3};a^{3};a^{3};a^{3};a^{3};a^{3};a^{3};a^{2};a^{2};a^{2};a^{2};
\nonumber \\
&&a;a;a;a;a;a\}  \label{EE88}
\end{eqnarray}
\noindent

\subsection{Solution for the \textbf{g\TEXTsymbol{\vert}\~{g}}--theories}

As already indicated in section 3, when introducing the variables $%
x_{a}^{i}=\prod_{b=1}^{\ell }(Q_{b}^{i})^{-K_{ab}}$ the constant
TBA-equations (\ref{cTBA}), or equivalently (\ref{Qr}) at certain fixed
points, acquire the more symmetric form 
\begin{equation}
\left( Q_{a}^{i}\right) ^{2}=\prod_{b=1}^{\ell }\left( Q_{b}^{i}\right)
^{I_{ab}}+\prod_{j=1}^{\tilde{\ell}}\left( Q_{a}^{j}\right) ^{\tilde{I}%
_{ij}}\,\,.  \label{QQ}
\end{equation}
It is convenient to take here $Q_{a}^{0}=Q_{0}^{i}=1$. We will now identify
the $Q$'s with various combinations of Weyl characters (\ref{QQ}) \ either
of the algebra \textbf{g} or \textbf{\~{g}} such that the relations (\ref{QQ}%
) are solved. One should note here that in (\ref{QQ}) the two algebras are
on the same footing, despite the fact that on the level of the scattering
matrix, i.e. the data which enter the Virasoro characters (\ref{chi}) and in
(\ref{cqr}) they play quite distinct roles. We always choose 
\begin{equation}
\tau =\frac{1}{h+\tilde{h}}
\end{equation}
in (\ref{ww}), with $h,\tilde{h}$ being the Coxeter numbers of \textbf{g}, 
\textbf{\~{g}}, respectively. It will be sufficient to concentrate on the 
\textbf{g}\TEXTsymbol{\vert}\textbf{\~{g}}-theories, since the coset models
reported on in section 3.4 may be constructed simply by means of a system of
the type (\ref{cTBA}). Having \ solved (\ref{QQ}) we also compute the
(effective) central charge according to (\ref{cqr}). In many cases this can
be done analytically by reducing the expression to some well-known (see e.g. 
\cite{dilog}) numerical relations for Rogers dilogarithm, such as $\mathcal{L%
}\left( 1/2\right) =\pi ^{2}/12$, $\mathcal{L}\left( (\sqrt{5}-1)/2\right)
=\pi ^{2}/10$, etc., by a successive application of the five term relation 
\begin{equation}
\mathcal{L}\left( x\right) +\mathcal{L}\left( y\right) =\mathcal{L}\left(
xy\right) +\mathcal{L}\left( \frac{x(1-y)}{1-xy}\right) +\mathcal{L}\left( 
\frac{y(1-x)}{1-xy}\right) \,\,.
\end{equation}
In several cases we do not attempt to be entirely rigorous and only verify
the relations numerically. Especially when a generic rank is involved we
only compute a large part of the beginning of the sequence and do not
attempt to perform inductive proofs.

We proceed case-by-case.

\subsubsection{\textbf{A}$_{\ell }$\TEXTsymbol{\vert}\textbf{A}$_{\tilde{\ell%
}}$}

In this case the recurrence relations (\ref{QQ}) are explicitly 
\begin{equation}
\left( Q_{a}^{k}\right) ^{2}=Q_{a+1}^{k}Q_{a-1}^{k}+Q_{a}^{k+1}Q_{a}^{k-1}
\label{QAA}
\end{equation}
for $1\leq a\leq \ell $, $1\leq k\leq \tilde{\ell}$. As was first pointed
out in \cite{Resh}, by identifying the Q's with Weyl characters these
relations may be solved explicitly. We may either use the characters $\chi
,\,\tilde{\chi}$ of \textbf{A}$_{\ell }$, \textbf{A}$_{\tilde{\ell}}$,
respectively, with $\tau =1/(\ell +\tilde{\ell}+2)$ 
\begin{equation}
Q_{a}^{k}=\chi _{k\lambda _{a}}(\tau )=\tilde{\chi}_{a\lambda _{k}}(\tau ).
\label{AA}
\end{equation}
This follows now immediately by noting that (\ref{QAA}) coincides precisely
with equation (\ref{h3}). Using these solutions, the central charges
according to (\ref{cqr}) turn out to be 
\begin{equation}
c=\frac{6}{\pi ^{2}}\sum\limits_{a=1}^{\ell }\sum\limits_{k=1}^{\tilde{\ell}}%
\mathcal{L}\left( \frac{\tilde{\chi}_{a\lambda _{k-1}}(\tau )\tilde{\chi}%
_{a\lambda _{k+1}}(\tau )}{\tilde{\chi}_{a\lambda _{k}}(\tau )^{2}}\right)
\,=\frac{\ell \tilde{\ell}(\tilde{\ell}+1)}{\ell +\tilde{\ell}+2}\,.
\end{equation}

\subsubsection{\textbf{A}$_{1}|$\textbf{\~{g}}--theories}

For the reasons mentioned in the previous section these particular
HSG-models are interesting to investigate. Exploiting the symmetry in the
equations (\ref{QQ}), they may be solved by appealing to the solutions which
correspond to the ones of minimal affine Toda field theories, i.e. \textbf{g%
\TEXTsymbol{\vert}A}$_{1}$. These solutions in terms of the characters of 
\textbf{g }may be extracted from the general formulae provided in \cite
{Resh,Kun}. The corresponding values were also stated thereafter in the
first reference in \cite{TBAKM} without proof. We demonstrate that
alternatively one may simply use combinations of the characters of \textbf{A}%
$_{1}$ 
\begin{equation}
\chi _{k\lambda }(\tau )=\frac{\sin (\pi (1+k)\tau )}{\sin (\pi \tau )}
\label{A1}
\end{equation}
in order to solve the recurrence relations.

\paragraph{\textbf{A}$_{1}|$\textbf{A}$_{\tilde{\ell}}$}

As a special case of (\ref{AA}) we obtain 
\begin{equation}
Q^{i}=\tilde{\chi}_{\lambda _{i}}(1/(\tilde{\ell}+3))\,=\chi _{i\lambda }(1/(%
\tilde{\ell}+3))\,.
\end{equation}
Translating to the $x$-variables we recover the values quoted in \cite{TBAKM}%
. The particularization of (\ref{cqr}) yields the central charges 
\begin{equation}
c=\frac{6}{\pi ^{2}}\sum_{k=1}^{\tilde{\ell}}\mathcal{L}\left( 1-\tilde{\chi}%
_{\lambda _{k}}(1/(\tilde{\ell}+3))\right) =\frac{\tilde{\ell}(\tilde{\ell}%
+1)}{\tilde{\ell}+3}\,\,.
\end{equation}

\paragraph{\textbf{A}$_{1}|$\textbf{D}$_{\tilde{\ell}}$}

We may express the solutions for this case either in terms of the \textbf{D}$%
_{\tilde{\ell}}$- or the \textbf{A}$_{1}$-Weyl characters. Taking $\tau =1/2%
\tilde{\ell}$ we obtain 
\begin{eqnarray}
Q^{i} &=&\sum\limits_{k=1}^{\frac{i+1}{2}}\tilde{\chi}_{\lambda
_{2k-1}}(\tau )=(i+1)\chi _{\lambda _{2\tilde{\ell}-2}}(\tau )=i+1\qquad
\qquad i\,\,\text{odd, }i\leq \tilde{\ell}-2,  \label{DA1} \\
Q^{i} &=&1+\sum\limits_{k=1}^{\frac{i}{2}}\tilde{\chi}_{\lambda _{2k}}(\tau
)=(i+1)\chi _{\lambda _{2\tilde{\ell}-2}}(\tau )=i+1\qquad \quad i\,\,\text{%
even, }i\leq \tilde{\ell}-2,  \label{DA2} \\
Q^{\tilde{\ell}-1} &=&\tilde{\chi}_{\lambda _{\tilde{\ell}-1}}(\tau )=Q^{%
\tilde{\ell}}=\tilde{\chi}_{\lambda _{\tilde{\ell}}}(\tau )=\sqrt{\tilde{\ell%
}}\,=\prod_{k=1}^{\tilde{\ell}-1}\chi _{(2k-1)\lambda }/\chi _{(\tilde{\ell}%
+k-1)\lambda }\,\,.  \label{DA3}
\end{eqnarray}
From the explicit expressions in section 4.1.2. follows that $\tilde{\chi}%
_{\lambda k}(\tau )=2$ for $k\leq \tilde{\ell}-2$ and the last relation in (%
\ref{DA3}). Therefore we may trivially evaluate the sums in (\ref{DA1}) and (%
\ref{DA2}), whose result we can employ to convince ourselves that (\ref{QQ})
is indeed satisfied. Once again translating to the $x$-variables yields the
values quoted in \cite{TBAKM}. According to (\ref{cqr}) the central charges
are then computed to 
\begin{equation}
c=\frac{6}{\pi ^{2}}\left( \sum_{k=1}^{\tilde{\ell}-3}\mathcal{L}\left( 
\frac{k(k+2)}{(k+1)^{2}}\right) +\mathcal{L}\left( \frac{\tilde{\ell}(\tilde{%
\ell}-2)}{(\tilde{\ell}-1)^{2}}\right) +2\mathcal{L}\left( 1-\tilde{\ell}%
^{-1}\right) \right) =\tilde{\ell}-1\,\,.
\end{equation}

\paragraph{$\mathbf{A}_{1}\mathbf{|E}_{6}$}

Using the conventions of section 4.1.3. the recurrence relations (\ref{QQ})
read in this case 
\begin{equation}
(Q^{1})^{2}=1+Q^{3},\quad (Q^{2})^{2}=1+Q^{4},\quad
(Q^{3})^{2}=1+Q^{4}Q^{1},\,\,\,\,\,(Q^{4})^{2}=1+Q^{2}(Q^{3})^{2},
\label{EQ2}
\end{equation}

\noindent where we have already exploited $Q^{1}=Q^{6}$, $Q^{3}=Q^{5}$,
which is a consequence of the symmetry of the Dynkin diagram. For $a=1$ and $%
\tau =1/14$ the expressions (\ref{E1})-(\ref{E4}) for the $E_{6}$-characters
reduce to 
\begin{equation}
\tilde{\chi}_{\lambda _{1}}=(2\,\sin \frac{\pi }{14})^{-1},\quad \tilde{\chi}%
_{\lambda _{2}}=\tilde{\chi}_{\lambda _{3}}=2\,\mathnormal{\cos }{\frac{\pi 
}{7},\quad }\tilde{\chi}_{\lambda _{4}}=0\,,
\end{equation}
such that we can identify them with combinations of $A_{1}$-characters and
vice versa

\begin{equation}
\tilde{\chi}_{\lambda _{1}}=1+\chi _{4\lambda }-\chi _{2\lambda },\quad
\qquad \tilde{\chi}_{\lambda _{2}}=\tilde{\chi}_{\lambda _{3}}=\chi
_{2\lambda }-1{\,\,}.  \label{rr}
\end{equation}

\noindent With these simple expressions for the characters, we may easily
check that the expressions 
\begin{equation}
Q^{1}=1+\chi _{4\lambda }-\chi _{2\lambda },\,\,\,\,\,Q^{2}=\chi _{2\lambda
},\,\,\,\,\,Q^{3}=\chi _{4\lambda },\,\,\,\,\,Q^{4}=\chi _{4\lambda }+\chi
_{2\lambda },
\end{equation}

\noindent indeed satisfy the relations (\ref{EQ2}). Of course with the help
of (\ref{rr}) it is also possible to express the $Q$'s in terms of the $%
\tilde{\chi}$'s instead of the $\chi $'s. Making then use of the symmetry
between the two algebras in (\ref{QQ}) and translating to the $x$-variables
we recover the numerical values quoted in \cite{TBAKM}. Assembling this, the
central charge according to (\ref{cqr}) is computed to 
\begin{equation}
c=\frac{6}{\pi ^{2}}\left( 2\mathcal{L}\left( \frac{Q^{3}}{(Q^{1})^{2}}%
\right) +\mathcal{L}\left( \frac{Q^{4}}{(Q^{2})^{2}}\right) +2\mathcal{L}%
\left( \frac{Q^{1}Q^{4}}{(Q^{3})^{2}}\right) +\mathcal{L}\left( \frac{%
(Q^{3})^{2}Q^{2}}{(Q^{4})^{2}}\right) \right) =\frac{36}{7}\,\,.
\end{equation}

\paragraph{$\mathbf{A}_{1}\mathbf{|E}_{7}$}

With the conventions of section 4.1.4. the recurrence relations (\ref{QQ})
read in this case 
\begin{eqnarray}
(Q^{1})^{2}
&=&1+Q^{3},\,\,\,\,\,\,\,\,\,(Q^{2})^{2}=1+Q^{4},\,\,\,\,%
\,(Q^{3})^{2}=1+Q^{4}Q^{1},\,\,\,\,\,(Q^{4})^{2}=1+Q^{3}Q^{5}Q^{2},\quad
\,\,\,  \label{Q71} \\
(Q^{5})^{2}
&=&1+Q^{4}Q^{6},\,\,(Q^{6})^{2}=1+Q^{7}Q^{5},\,\,\,\,\,(Q^{7})^{2}=1+Q^{6}.
\label{Q72}
\end{eqnarray}

\noindent For $a=1$ and $\tau =1/20$ the expressions (\ref{E11})-(\ref{E77})
for the $E_{7}$-characters simplify to 
\begin{equation}
\tilde{\chi}_{\lambda _{1}}=\tilde{\chi}_{\lambda _{6}}=\frac{\sin \frac{%
3\pi }{5}}{\sin \frac{\pi }{5}}\quad \tilde{\chi}_{\lambda _{2}}=\sqrt{2}%
,\quad \tilde{\chi}_{\lambda _{3}}=\tilde{\chi}_{\lambda _{4}}=\tilde{\chi}%
_{\lambda _{5}}=0,\quad \tilde{\chi}_{\lambda _{7}}=\frac{\sqrt{2}}{4\sin 
\frac{\pi }{20}\sin \frac{9\pi }{20}},
\end{equation}
such that by recalling (\ref{A1}) we can identify them with combinations of $%
A_{1}$-characters and vice versa 
\begin{equation}
\tilde{\chi}_{\lambda _{1}}=\tilde{\chi}_{\lambda _{6}}=\chi _{4\lambda
}-\chi _{2\lambda },\quad \quad \tilde{\chi}_{\lambda _{2}}=\chi _{5\lambda
}-\chi _{3\lambda },\quad \,\,\quad \,\tilde{\chi}_{\lambda _{7}}=\chi
_{9\lambda }+\chi _{\lambda }-\chi _{7\lambda }.
\end{equation}

\noindent With these simple expressions for the characters, we may once
again verify after exploiting the symmetry of (\ref{QQ}) or by direct
analysis with the $\mathbf{A}_{1}$-characters, that the expressions proposed
in \cite{Kun} 
\begin{eqnarray}
Q^{1} &=&1+\tilde{\chi}_{\lambda _{1}},\,\,\,\,\quad \quad \quad \,\,\,Q^{2}=%
\tilde{\chi}_{\lambda _{7}}+\tilde{\chi}_{\lambda _{2}},\quad Q^{3}=1+3%
\tilde{\chi}_{\lambda _{1}},\quad Q^{4}=3+6\tilde{\chi}_{\lambda _{1}},
\label{Q777} \\
\,Q^{5} &=&2\tilde{\chi}_{\lambda _{7}}+2\tilde{\chi}_{\lambda _{2}},\quad
\quad Q^{6}=1+2\tilde{\chi}_{\lambda _{1}},\,\,\quad \,\,Q^{7}=\tilde{\chi}%
_{\lambda _{7}},  \label{Q7777}
\end{eqnarray}
indeed satisfy (\ref{Q71})-(\ref{Q72})\footnote{%
There appears to be a small typo in Eq. (A.11.c) of \cite{Kun}, which reads
when translated to our conventions, i.e. $6\rightarrow 7$, $Q_{7}=\chi
_{\lambda _{1}}$instead of (\ref{Q7777}).}. Renaming our roots and
translating to the $x$-variables, we recover the numerical values quoted in 
\cite{TBAKM}. The central charge (\ref{cqr}) is in this case 
\begin{eqnarray}
c &=&\frac{6}{\pi ^{2}}\left( \mathcal{L}\left( \frac{3\sqrt{5}-5}{2}\right)
+\mathcal{L}\left( \frac{3\sqrt{5}-3)}{4}\right) +\mathcal{L}\left( \frac{3%
\sqrt{5}+3)}{10}\right) +\mathcal{L}\left( \frac{4\sqrt{5}}{9}\right) \right.
\nonumber \\
&&\left. +\mathcal{L}\left( \frac{3(3+\sqrt{5})}{16}\right) +\mathcal{L}%
\left( \frac{1+\sqrt{5}}{4}\right) +\mathcal{L}\left( 4(\sqrt{5}-4)\right)
\right) =\frac{63}{10}\,\,.
\end{eqnarray}

\paragraph{$\mathbf{A}_{1}\mathbf{|E}_{8}$}

The recurrence relations (\ref{QQ}) read now 
\begin{eqnarray}
(Q^{1})^{2}
&=&1+Q^{3},\,\,\,(Q^{2})^{2}=1+Q^{4},\,\,\,(\,Q^{3})^{2}=1+Q^{1}Q^{4},\,%
\,(Q^{4})^{2}=1+Q^{3}Q^{2}Q^{5},\,\,\,\,\,\,\,\,\,\,  \label{bb} \\
\,(Q^{5})^{2}
&=&1+Q^{4}Q^{6},\,\,\,(\,Q^{6})^{2}=1+Q^{5}Q^{7},\,\,%
\,(Q^{7})^{2}=1+Q^{6}Q^{8},\,\,\,(Q^{8})^{2}=1+Q^{7}.  \label{bbb}
\end{eqnarray}
\noindent When setting $a=1$ and $\tau =1/32$, the $E_{8}$-characters (\ref
{EE81})-(\ref{EE88}) reduce to

\begin{equation}
\tilde{\chi}_{\lambda _{1}}=1,\,\,\,\,\,\,\,\,\tilde{\chi}_{\lambda _{8}}=%
\sqrt{2},\quad \tilde{\chi}_{\lambda _{2}}=\tilde{\chi}_{\lambda _{3}}=%
\tilde{\chi}_{\lambda _{4}}=\tilde{\chi}_{\lambda _{5}}=\tilde{\chi}%
_{\lambda _{6}}=\tilde{\chi}_{\lambda _{7}}=0\,\,.
\end{equation}

\noindent We may then identify them with combinations of $A_{1}$-characters 
\begin{equation}
\,\tilde{\chi}_{\lambda _{1}}=\chi _{30\lambda },\qquad \tilde{\chi}%
_{\lambda _{8}}=\chi _{8\lambda }-\chi _{\lambda }\,.\quad
\end{equation}

\noindent With these numerical values we can express the solutions of \ (\ref
{bb}) and \ (\ref{bbb}) in terms of the $E_{8}/A_{1}$-characters 
\begin{eqnarray}
Q^{1} &=&2+\tilde{\chi}_{\lambda _{8}},\quad \quad \,\,Q^{2}=3+2\tilde{\chi}%
_{\lambda _{8}},\,\,\,\,\,\,Q^{3}=5+4\tilde{\chi}_{\lambda
_{8}},\,\,\,\,\,\,\,Q^{4}=4(4+3\tilde{\chi}_{\lambda _{8}}), \\
Q^{5} &=&3(3+2\tilde{\chi}_{\lambda _{8}}),\,\,\,\,Q^{6}=5+3\tilde{\chi}%
_{\lambda _{8}},\quad Q^{7}=2+2\tilde{\chi}_{\lambda _{8}},\quad Q^{8}=%
\tilde{\chi}_{\lambda _{1}}+\tilde{\chi}_{\lambda _{8}}\,\,.
\end{eqnarray}

\noindent In \cite{Kun} only the values for $Q^{1}$ and $Q^{8}$ were
presented. As in the previous case, after relabeling our roots and
translating to the $x$-variables we recover the numbers quoted in \cite
{TBAKM}. In this case the central charge (\ref{cqr}) equals 
\begin{eqnarray}
c &=&\frac{6}{\pi ^{2}}\left( \mathcal{L}\left( \sqrt{2}-\frac{1}{2}\right) +%
\mathcal{L}\left( 12\sqrt{2}-16\right) +\mathcal{L}\left( \frac{40\sqrt{2}-8%
}{49}\right) +\mathcal{L}\left( \frac{12\sqrt{2}+15}{32}\right) \right. 
\nonumber \\
&&\left. +\mathcal{L}\left( \frac{12\sqrt{2}-8}{9}\right) +\mathcal{L}\left( 
\frac{30\sqrt{2}+6}{49}\right) +\mathcal{L}\left( \frac{1}{4}+\frac{1}{\sqrt{%
2}}\right) +\mathcal{L}\left( 2\sqrt{2}-2\right) \right) =\frac{15}{2}%
.\,\,\,\,\,\,\,\,\,\,\,
\end{eqnarray}

\subsubsection{\textbf{D}$_{\ell }$\TEXTsymbol{\vert}\textbf{A}$_{\tilde{\ell%
}}$}

In this case the recurrence relations (\ref{QQ}) read 
\begin{eqnarray}
\left( Q_{a}^{k}\right) ^{2}
&=&Q_{a+1}^{k}Q_{a-1}^{k}+Q_{a}^{k+1}Q_{a}^{k-1},\quad \quad 1\leq a\leq
\ell -3 \\
\left( Q_{\ell -2}^{k}\right) ^{2} &=&Q_{\ell -2}^{k-1}Q_{\ell
-2}^{k+1}+Q_{\ell }^{k}Q_{\ell -1}^{k}Q_{\ell -3}^{k} \\
\left( Q_{p}^{k}\right) ^{2} &=&Q_{\ell -2}^{k}+Q_{p}^{k+1}Q_{p}^{k-1},\quad
\quad \quad p=\ell ,\ell -1
\end{eqnarray}
for $1\leq k\leq \tilde{\ell}$. Also in this case we may exploit the
symmetry of equations (\ref{QQ}) in the two algebras. We simply have to
exchange their roles in order to obtain a solution for the \textbf{D}$_{\ell
}$\TEXTsymbol{\vert}\textbf{A}$_{\tilde{\ell}}$-theory from the one for the 
\textbf{A}$_{\tilde{\ell}}$\TEXTsymbol{\vert}\textbf{D}$_{\ell }$ reported
in \cite{Resh,Kun}. Taking $\tau =1/(2\ell +\tilde{\ell}-1)$, we can
express, following \cite{Resh,Kun}, the $Q$'s in terms of the Weyl
characters of \textbf{D}$_{\ell }$. 
\begin{eqnarray}
Q_{s}^{k} &=&\sum\limits_{l_{1}=0}^{k}\ldots \sum\limits_{l_{s-2}=0}^{k}\chi
_{k\lambda _{s}+l_{1}(\lambda _{1}-\lambda _{s})+\ldots +l_{s-2}(\lambda
_{s-2}-\lambda _{s})}(\tau ),\,\, \\
Q_{p}^{k} &=&\!\!\!\sum\limits_{\tilde{a}=0}^{k}\sum\limits_{l_{2}=0}^{%
\tilde{a}}\ldots \!\!\!\sum\limits_{l_{p-2}=0}^{\tilde{a}}\chi _{\tilde{a}%
\lambda _{p}+l_{2}(\lambda _{2}-\lambda _{p})+\ldots +l_{p-2}(\lambda
_{p-2}-\lambda _{p})}(\tau ), \\
Q_{\ell -1}^{k} &=&\chi _{k\lambda _{\ell -1}}(\tau ),\qquad Q_{\ell
}^{k}=\chi _{k\lambda _{\ell }}(\tau \,)\,.
\end{eqnarray}
Here $s$ and $p$ are odd and even integers smaller $\ell -1$, respectively.
Alternatively we may also express the $Q$'s in terms of the $A_{\tilde{\ell}%
} $-characters. For instance for $\mathbf{D}_{\ell }\mathbf{|A}_{2}\mathbf{\,%
}$\ we find 
\begin{eqnarray}
Q_{2k}^{1} &=&Q_{2k}^{2}=1+\sum\limits_{i=1}^{k}\left( \tilde{\chi}%
_{i\lambda }-\tilde{\chi}_{(i-2)\lambda }+\tilde{\chi}_{(\ell -i)\lambda }-%
\tilde{\chi}_{(\ell -i-2)\lambda }\right) ,\,\,\quad 2k<\ell -1,\,\,\,\,\, \\
Q_{2k-1}^{1} &=&Q_{2k-1}^{2}=\sum\limits_{i=0}^{k-1}\left( \tilde{\chi}%
_{i\lambda }-\tilde{\chi}_{(i-2)\lambda }\right)
+\sum\limits_{i=1}^{k}\left( \tilde{\chi}_{(\ell -i)\lambda }-\tilde{\chi}%
_{(\ell -i-2)\lambda }\right) ,\,\,\quad 2k<\ell ,\,\,\,\,\, \\
Q_{\ell }^{1} &=&Q_{\ell -1}^{1}=Q_{\ell }^{2}=Q_{\ell -1}^{2}=\tilde{\chi}%
_{\ell \lambda }-\tilde{\chi}_{(\ell -2)\lambda }\,.
\end{eqnarray}
We suppressed the $\tau $-dependence, denote $\lambda =\lambda _{1}=\lambda
_{2}$ and recall that we take $\tilde{\chi}_{i\lambda }=1$ for $i=0$, $%
\tilde{\chi}_{i\lambda }=0$ for $i<0$.

Let us now consider some theories which may not be obtained from others
previously studied, by exploiting the symmetry properties of the recurrence
relations (\ref{QQ}).

\subsubsection{\textbf{D}$_{\ell }$\TEXTsymbol{\vert}\textbf{D}$_{\tilde{\ell%
}}$}

The recurrence relations (\ref{QQ}) are now constructed from the symmetric $%
D_{l}$-incidence matrix, whose non-vanishing entries are 
\begin{equation}
\hat{I}_{t,t+1}=1\quad 1\leq t\leq l-2,\quad \quad \hat{I}_{t,t-1}=1\quad
2\leq t\leq l-1,\quad \quad \hat{I}_{l,l-2}=1\,,
\end{equation}
such that $I=\hat{I}$ with $l=\ell $ and $\tilde{I}=\hat{I}$ with $l=\tilde{%
\ell}$.

\paragraph{\textbf{D}$_{4}$\TEXTsymbol{\vert}\textbf{D}$_{4}$}

For the choice $\tau =1/12$ the $D_{4}$-characters (\ref{11}) and (\ref{22})
become 
\begin{eqnarray}
\chi _{\lambda _{1}} &=&3+\sqrt{3},\quad \quad \chi _{2\lambda _{1}}=5+3%
\sqrt{3},\quad \chi _{3\lambda _{1}}=6+4\sqrt{3},\quad \chi _{\lambda
_{2}}=6+3\sqrt{3}, \\
\chi _{2\lambda _{2}} &=&15+9\sqrt{3},\quad \chi _{3\lambda _{2}}=10+6\sqrt{3%
},\quad \chi _{\lambda _{3}}=\chi _{\lambda _{1}},\,\,\,\,\chi _{2\lambda
_{3}}=\chi _{2\lambda _{1}},\,\,\,\,\chi _{3\lambda _{3}}=\chi _{3\lambda
_{1}}.\,\,\,\,\,\,\,\,\,\,
\end{eqnarray}
\noindent The recurrence relations (\ref{QQ}) are solved by 
\begin{eqnarray}
Q_{1}^{1}
&=&Q_{1}^{3}=Q_{1}^{4}=Q_{3}^{1}=Q_{4}^{1}=Q_{3}^{3}=Q_{4}^{3}=Q_{3}^{4}=Q_{4}^{4}=4\chi _{\lambda _{1}}-\chi _{3\lambda _{1}}=6,
\\
Q_{1}^{2} &=&Q_{3}^{2}=Q_{4}^{2}=Q_{2}^{3}=Q_{2}^{1}=Q_{2}^{4}=18,\qquad
\qquad Q_{2}^{2}=108.
\end{eqnarray}
\noindent

\noindent The central charge (\ref{cqr}) is in this case simply 
\begin{equation}
c=\frac{6}{\pi ^{2}}\left( 10\mathcal{L}\left( \frac{1}{2}\right) +3\mathcal{%
L}\left( \frac{2}{3}\right) +3\mathcal{L}\left( \frac{1}{3}\right) \right)
=8\,\,.
\end{equation}

\paragraph{$\mathbf{D}_{4}\mathbf{|D}_{5}$}

We take now $\tau =1/14$ such that some of the $D_{4}$-characters (\ref{11})
and (\ref{22}) read, 
\begin{eqnarray}
\chi _{\lambda _{1}} &=&\frac{\sin \frac{2\pi }{7}\sin \frac{3\pi }{7}}{\sin 
\frac{\pi }{14}\sin \frac{3\pi }{14}},\quad \chi _{2\lambda _{1}}=2\chi
_{\lambda _{1}}\cos \frac{\pi }{7}\sin \frac{5\pi }{14}=\chi _{4\lambda
_{1}}/2,\quad \chi _{3\lambda _{1}}=\frac{\chi _{4\lambda _{1}}\cos ^{2}{{%
\frac{3\pi }{7}}}}{\sin {\frac{5\pi }{14}}},\quad \,\,\, \\
\chi _{\lambda _{2}} &=&\chi _{2\lambda _{1}}=\chi _{4\lambda _{2}}/4,\quad
\chi _{2\lambda _{2}}=\frac{\chi _{2\lambda _{1}}\sin ^{2}{\frac{5\pi }{14}%
\sin {\frac{2\pi }{7}}}}{\sin {\frac{3\pi }{14}}\sin ^{2}{\frac{\pi }{7}}}%
,\quad \chi _{3\lambda _{2}}=\frac{\chi _{2\lambda _{2}}\sin ^{2}{\frac{3\pi 
}{7}}}{\sin {\frac{5\pi }{14}}\sin ^{2}{\frac{2\pi }{7}}},
\end{eqnarray}
and the ones for $D_{5}$%
\begin{eqnarray}
\tilde{\chi}_{\lambda _{1}} &=&\,\frac{\sin \frac{5\pi }{14}\sin \frac{3\pi 
}{7}}{\sin \frac{\pi }{14}\sin \frac{2\pi }{7}},\quad \quad \tilde{\chi}%
_{2\lambda _{1}}\,=\tilde{\chi}_{\lambda _{1}}\frac{\sin \frac{3\pi }{7}}{%
\sin \frac{\pi }{7}},\quad \tilde{\chi}_{3\lambda _{1}}=\chi _{2\lambda
_{1}}, \\
\tilde{\chi}_{\lambda _{2}} &=&\tilde{\chi}_{\lambda _{1}}\frac{\sin \frac{%
5\pi }{14}\sin \frac{2\pi }{7}}{\sin \frac{3\pi }{14}\sin \frac{\pi }{7}}%
,\quad \tilde{\chi}_{2\lambda _{2}}=2\tilde{\chi}_{2\lambda _{1}}\frac{\cos 
\frac{3\pi }{14}}{\sin \frac{\pi }{7}},\quad \tilde{\chi}_{3\lambda _{2}}=(%
\tilde{\chi}_{\lambda _{2}})^{2}/\tilde{\chi}_{\lambda _{1}}, \\
\tilde{\chi}_{\lambda _{3}} &=&2\tilde{\chi}_{\lambda _{2}}\cos \frac{\pi }{7%
},\qquad \tilde{\chi}_{2\lambda _{3}}=\tilde{\chi}_{3\lambda _{2}}/2,\qquad 
\tilde{\chi}_{3\lambda _{3}}=\tilde{\chi}_{2\lambda _{1}}/2, \\
\tilde{\chi}_{\lambda _{4}} &=&\frac{\sin ^{2}\frac{3\pi }{7}}{\sin \frac{%
\pi }{14}\sin \frac{3\pi }{14}},\quad \tilde{\chi}_{2\lambda _{4}}=\tilde{%
\chi}_{2\lambda _{1}}\frac{1}{\sin (\frac{3\pi }{14})},\quad \tilde{\chi}%
_{3\lambda _{4}}=\chi _{3\lambda _{1}}\,.
\end{eqnarray}
We may then express the characters of $D_{5}$ in terms of characters of $%
D_{4}$ 
\begin{eqnarray}
\tilde{\chi}_{\lambda _{1}} &=&({\chi }_{3\lambda _{1}}-{\chi }_{\lambda
_{2}})/2,\quad \tilde{\chi}_{2\lambda _{1}}=(\chi _{3\lambda
_{1}}-2)/2,\quad \tilde{\chi}_{3\lambda _{1}}=\chi _{2\lambda _{1}}, \\
\tilde{\chi}_{\lambda _{2}} &=&(\chi _{3\lambda _{1}}+\chi _{\lambda
_{2}}-2\chi _{\lambda _{1}}-2)/2,\quad \tilde{\chi}_{2\lambda _{2}}=(\chi
_{3\lambda _{2}}-\chi _{3\lambda _{1}})/2,\quad \\
\tilde{\chi}_{3\lambda _{2}} &=&(-10\chi _{\lambda _{1}}+9(\chi _{2\lambda
_{1}}-1)+6\chi _{3\lambda _{1}}-\chi _{2\lambda _{2}})/2,\quad \\
\tilde{\chi}_{\lambda _{3}} &=&\chi _{3\lambda _{1}}-1,\quad \tilde{\chi}%
_{2\lambda _{3}}=2\chi _{4\lambda _{1}},\quad \tilde{\chi}_{3\lambda
_{3}}=(\chi _{3\lambda _{1}}-2)/2, \\
\tilde{\chi}_{\lambda _{4}} &=&(\chi _{3\lambda _{1}}-2\chi _{\lambda
_{1}})/2,\quad \tilde{\chi}_{2\lambda _{4}}=\chi _{3\lambda _{1}}-\chi
_{\lambda _{1}}-1,\quad \tilde{\chi}_{3\lambda _{4}}=\chi _{3\lambda _{1}}.
\end{eqnarray}
In terms of these quantities we may then solve the recurrence relations by 
\begin{eqnarray}
Q_{1}^{1} &=&1+{\chi }_{\lambda _{1}}, \\
Q_{1}^{2} &=&6(\chi _{3\lambda _{1}}+\chi _{3\lambda _{2}}+1)-10(\chi
_{\lambda _{1}}+\chi _{2\lambda _{1}}+\chi _{4\lambda _{1}})+4\chi
_{2\lambda _{2}}-9\chi _{4\lambda _{2}}, \\
Q_{1}^{3} &=&2(2-\chi _{\lambda _{1}}+\chi _{2\lambda _{1}}+\chi _{2\lambda
_{2}}-\chi _{3\lambda _{2}}+\chi _{4\lambda _{2}}), \\
Q_{1}^{4} &=&10(\chi _{2\lambda _{2}}-\chi _{3\lambda _{1}}-\chi _{4\lambda
_{1}}-\chi _{4\lambda _{2}})-8\chi _{\lambda _{1}}-5\chi _{2\lambda
_{1}}+6\chi _{3\lambda _{2}}-7, \\
Q_{2}^{1} &=&8(\chi _{2\lambda _{2}}-\chi _{\lambda _{1}}-\chi _{4\lambda
_{1}}-\chi _{4\lambda _{2}}+1)+5(\chi _{3\lambda _{1}}-\chi _{2\lambda
_{1}})+2\chi _{3\lambda _{2}}, \\
Q_{2}^{2} &=&8(\chi _{3\lambda _{2}}+\chi _{3\lambda _{1}}-\chi _{\lambda
_{1}}-\chi _{4\lambda _{2}})-5\chi _{4\lambda _{1}}-4\chi _{2\lambda _{2}}+2,
\\
Q_{2}^{3} &=&6(\chi _{\lambda _{1}}+\chi _{4\lambda _{1}}+\chi _{4\lambda
_{2}})+4\chi _{2\lambda _{1}}-2\chi _{3\lambda _{1}}+1, \\
Q_{2}^{4} &=&Q_{2}^{5}=6(\chi _{4\lambda _{2}}-\chi _{2\lambda _{2}})+4(\chi
_{\lambda _{1}}+\chi _{4\lambda _{1}})-\chi _{3\lambda _{2}}+\chi _{3\lambda
_{1}}, \\
Q_{3}^{1}
&=&Q_{4}^{1}=Q_{1}^{1},\,\,\,\,\,\,\,Q_{3}^{2}=Q_{4}^{2}=Q_{1}^{2},\,\,\,\,%
\,\,\,Q_{3}^{4}=Q_{3}^{5}=Q_{1}^{4}=Q_{4}^{4}=Q_{4}^{5}, \\
Q_{3}^{3} &=&Q_{4}^{3}=Q_{1}^{3}.
\end{eqnarray}
Using these values we compute numerically the central charge to $c=80/7$.

\paragraph{\textbf{D}$_{5}$\TEXTsymbol{\vert}\textbf{D}$_{5}$}

For $\tau =1/16$ and $\ell =5$ the $D_{5}$-characters (\ref{11}) and (\ref
{22}) become 
\begin{eqnarray}
\chi _{\lambda _{1}} &=&\sqrt{2}\,\frac{\sin \frac{5\pi }{16}}{\sin \frac{%
\pi }{16}},\,\,\,\,\quad \chi _{2\lambda _{1}}=4+3\sqrt{2}+2\sqrt{10+7\sqrt{2%
}},\quad \\
\chi _{3\lambda _{1}} &=&8+5\sqrt{2}+\sqrt{2(58+41\sqrt{2})},\quad \chi
_{4\lambda _{1}}=2\chi _{2\lambda _{1}}, \\
\,\,\,\,\chi _{\lambda _{2}} &=&\chi _{2\lambda _{1}}+1,\quad \quad \chi
_{2\lambda _{2}}=22+17\sqrt{2}+2\sqrt{274+193\sqrt{2}}, \\
\chi _{3\lambda _{2}} &=&46+32\sqrt{2}+6\sqrt{116+82\sqrt{2}},\quad \chi
_{4\lambda _{2}}=4+6\chi _{2\lambda _{1}}, \\
\chi _{\lambda _{3}} &=&2+2\chi _{\lambda _{2}},\,\,\,\,\quad \chi
_{2\lambda _{3}}=61+41\sqrt{2}+6\sqrt{194+137\sqrt{2}}, \\
\chi _{3\lambda _{3}} &=&100+69\sqrt{2}+13\sqrt{116+82\sqrt{2}},\quad \chi
_{4\lambda _{3}}=\chi _{4\lambda _{2}}, \\
\chi _{\lambda _{4}} &=&2(1+\sqrt{2}+\sqrt{2+\sqrt{2}}),\quad \quad \chi
_{2\lambda _{4}}=\chi _{3\lambda _{1}},\,\,\,\,\chi _{3\lambda _{4}}=2\chi
_{3\lambda _{1}}, \\
\chi _{4\lambda _{4}} &=&18+14\sqrt{2}+6\sqrt{20+14\sqrt{2}}\,.
\end{eqnarray}

\noindent Noting the symmetry $Q_{a}^{i}=Q_{i}^{a}$, we may now express the
Q's in terms of $D_{5}$-characters 
\begin{eqnarray}
Q_{1}^{1} &=&2(\chi _{\lambda _{2}}-\chi _{\lambda _{1}}-\chi _{\lambda
_{4}}),\quad \\
Q_{1}^{2} &=&2(\chi _{2\lambda _{3}}+\chi _{\lambda _{4}}-\chi _{\lambda
_{1}}-\chi _{2\lambda _{1}}-\chi _{3\lambda _{1}}-\chi _{2\lambda
_{2}})-\chi _{4\lambda _{1}}-\chi _{3\lambda _{4}}-\chi _{4\lambda _{4}}, \\
Q_{1}^{3} &=&2(\chi _{3\lambda _{3}}-\chi _{\lambda _{1}}-\chi _{2\lambda
_{1}}-\chi _{3\lambda _{1}}-\chi _{4\lambda _{1}}-\chi _{2\lambda _{2}}-\chi
_{\lambda _{4}}-\chi _{4\lambda _{4}})  \nonumber \\
&&+\chi _{3\lambda _{4}}-\chi _{\lambda _{2}}-\chi _{2\lambda _{3}}\quad , \\
Q_{1}^{4} &=&Q_{1}^{5}=\chi _{4\lambda _{1}}+\chi _{3\lambda _{2}}+\chi
_{2\lambda _{3}}-\chi _{\lambda _{1}}-\chi _{2\lambda _{1}}-\chi _{\lambda
_{2}}-\chi _{3\lambda _{3}}-\chi _{\lambda _{4}}, \\
Q_{2}^{2} &=&2(\chi _{4\lambda _{1}}+\chi _{2\lambda _{2}}+\chi _{2\lambda
_{3}}+\chi _{\lambda _{4}}+\chi _{4\lambda _{4}}-\chi _{\lambda _{1}}-\chi
_{2\lambda _{1}}-\chi _{3\lambda _{1}}-\chi _{4\lambda _{2}}-\chi _{3\lambda
_{4}})  \nonumber \\
&&+\chi _{3\lambda _{2}}-\chi _{\lambda _{2}}-\chi _{3\lambda _{3}}, \\
Q_{2}^{3} &=&2(\chi _{3\lambda _{2}}+\chi _{3\lambda _{3}}+\chi _{3\lambda
_{4}}-\chi _{\lambda _{1}}-\chi _{2\lambda _{1}}-\chi _{4\lambda _{1}}-\chi
_{2\lambda _{2}}-\chi _{\lambda _{4}}-\chi _{4\lambda _{4}})+\chi _{2\lambda
_{3}}  \nonumber \\
&&-\chi _{3\lambda _{1}}-\chi _{\lambda _{2}}-1, \\
Q_{2}^{4} &=&Q_{2}^{5}=1+\chi _{\lambda _{1}}+\chi _{4\lambda _{1}}+\chi
_{3\lambda _{2}}+\chi _{2\lambda _{3}}+\chi _{4\lambda _{4}}-\chi _{2\lambda
_{2}}-\chi _{3\lambda _{3}}, \\
Q_{3}^{3} &=&8(\chi _{2\lambda _{3}}+\chi _{3\lambda _{3}}+\chi _{4\lambda
_{4}}-\chi _{\lambda _{1}}-\chi _{2\lambda _{1}}-\chi _{3\lambda _{1}}-\chi
_{4\lambda _{1}}-\chi _{\lambda _{2}}-\chi _{2\lambda _{2}}) \\
&&+7(\chi _{3\lambda _{2}}+\chi _{4\lambda _{2}}+\chi _{3\lambda
_{4}})-5\chi _{\lambda _{4}}+4, \\
Q_{3}^{4} &=&Q_{3}^{5}=\chi _{4\lambda _{1}}+\chi _{3\lambda _{2}}+\chi
_{2\lambda _{3}}-\chi _{2\lambda _{1}}-\chi _{\lambda _{2}}-\chi _{2\lambda
_{2}}-\chi _{4\lambda _{2}}-\chi _{4\lambda _{4}}, \\
Q_{5}^{5} &=&Q_{4}^{4}=Q_{4}^{5}=1+\chi _{\lambda _{1}}+\chi _{4\lambda
_{1}}+\chi _{\lambda _{2}}+\chi _{2\lambda _{2}}+\chi _{4\lambda _{4}}-\chi
_{3\lambda _{1}}-\chi _{3\lambda _{2}}.
\end{eqnarray}
\noindent Using these values we compute numerically the central charge to $%
c=25/2$.

\subsubsection{$D_{4}|E_{6}$}

In this case recurrence relations (\ref{QQ}) read 
\begin{eqnarray}
(Q_{1}^{1})^{2} &=&Q_{2}^{1}+Q_{1}^{2},\,\,\,\,\,\,\quad \quad
\,\,\,\,(Q_{1}^{2})^{2}=Q_{2}^{2}+Q_{1}^{4},\,\,\,\quad
\,\,\,(Q_{1}^{3})^{2}=Q_{2}^{3}+Q_{1}^{4}Q_{1}^{1},\quad   \label{q1} \\
(Q_{1}^{4})^{2} &=&Q_{2}^{4}+Q_{1}^{2}(Q_{1}^{3})^{2},\quad
\,\,(Q_{2}^{1})^{2}=Q_{4}^{1}+Q_{2}^{2},\,\,\,\,\quad
\,\,(Q_{2}^{2})^{2}=Q_{4}^{2}+Q_{2}^{4},\quad  \\
(Q_{2}^{3})^{2}
&=&Q_{4}^{3}+Q_{2}^{4}Q_{2}^{1},\,\,\,\,\,\,\,\,\,\,\,\,\,\,%
\,(Q_{2}^{4})^{2}=Q_{4}^{4}+Q_{2}^{2}(Q_{2}^{3})^{2}.  \label{q3}
\end{eqnarray}
\noindent We already took the relations 
\begin{equation}
Q_{a}^{1}=Q_{a}^{6},\,\,\,\,\,\,\,\,\,\,\,\,Q_{a}^{3}=Q_{a}^{5},\,\,%
\,Q_{1}^{i}=Q_{3}^{i}=Q_{4}^{i},\,\,\,\,\,\,\,\,\,\,1\leq a\leq 4,1\leq
i\leq 6
\end{equation}
\noindent into account which arise as a consequence of the symmetries of the 
$D_{4}$ and $E_{6}$ Dynkin diagrams. Taking now $\tau =1/18$, $\ell =4$ and $%
\tilde{\ell}=6$ the $D_{4}$-characters turn out to be 
\begin{eqnarray}
\chi _{\lambda _{1}} &=&\sqrt{3}\,\frac{\sin \frac{2\pi }{9}}{\sin \frac{\pi 
}{18}},\,\,\,\,\,\,\,\,\,\,\,\,\,\chi _{2\lambda _{1}}=\sqrt{3}\,\frac{\sin 
\frac{5\pi }{18}\sin \frac{7\pi }{18}}{\sin \frac{\pi }{18}\sin \frac{\pi }{9%
}},\,\,\,\,\,\,\,\,\,\,\,\,\,\chi _{3\lambda _{1}}=\sqrt{3}\,\chi _{2\lambda
_{1}}\,\frac{\sin \frac{4\pi }{9}}{\sin \frac{5\pi }{18}}, \\
\chi _{4\lambda _{1}} &=&\frac{2}{\sqrt{3}}\,\chi _{3\lambda _{1}}\,\frac{%
\sin \frac{7\pi }{18}}{\sin \frac{2\pi }{9}},\,\,\,\,\,\,\chi _{5\lambda
_{1}}=\chi _{4\lambda _{1}}\,\frac{\sin ^{2}\frac{4\pi }{9}}{\sin \frac{7\pi 
}{18}\sin \frac{5\pi }{18}},\,\,\,\,\,\,\,\chi _{6\lambda _{1}}=\frac{2}{%
\sqrt{3}}\,\chi _{5\lambda _{1}}\,\frac{\sin \frac{7\pi }{18}}{\sin \frac{%
4\pi }{9}}, \\
\chi _{\lambda _{2}} &=&\frac{1}{6}\,\chi _{3\lambda _{1}}\,\frac{\tan \frac{%
2\pi }{9}}{\sin \frac{\pi }{9}},\,\,\,\,\,\,\,\,\,\,\,\,\,\chi _{2\lambda
_{2}}=\frac{2}{3}\,\chi _{3\lambda _{2}}\,\frac{\sin ^{2}\frac{2\pi }{9}}{%
\sin ^{2}\frac{7\pi }{18}},\,\,\,\,\,\,\,\,\,\,\,\,\,\chi _{3\lambda _{2}}=%
\sqrt{3}\,\chi _{2\lambda _{1}}^{2}\frac{\sin \frac{\pi }{18}}{\sin \frac{%
\pi }{9}}, \\
\chi _{4\lambda _{2}} &=&\frac{\sqrt{3}}{2}\,\chi _{4\lambda _{1}}\,\chi
_{2\lambda _{1}}\,\frac{\sin \frac{\pi }{18}}{\sin \frac{\pi }{9}}%
,\,\,\,\,\,\,\,\chi _{5\lambda _{2}}=\frac{2}{3}\,\chi _{4\lambda _{2}}\,%
\frac{\sin ^{2}\frac{4\pi }{9}}{\sin ^{2}\frac{5\pi }{18}},\,\,\,\,\,\,\chi
_{6\lambda _{2}}=\frac{4}{\sqrt{3}}\,\chi _{5\lambda _{2}}\,\frac{\sin \frac{%
4\pi }{9}\sin \frac{\pi }{18}}{\sin ^{2}\frac{7\pi }{18}},\,\,\,\,\,\,\,\,\,%
\,\,\,\,\,\,
\end{eqnarray}
and the $E_{6}$-characters are 
\begin{eqnarray}
\tilde{\chi}_{\lambda _{1}} &=&\frac{\sqrt{3}}{2\sin {\frac{2\pi }{9}}\sin 
\frac{\pi }{18}},\,\,\,\,\,\,\,\,\,\,\tilde{\chi}_{\lambda _{2}}=\frac{4}{%
\sqrt{3}}\,\tilde{\chi}_{\lambda _{1}}\,{\sin \frac{5\pi }{18}\sin \frac{%
4\pi }{9}},\,\,\,\,\,\,\,\,\,\,\tilde{\chi}_{\lambda _{3}}=\frac{3}{4}\,%
\tilde{\chi}_{\lambda _{2}}\,\frac{\cos \frac{\pi }{9}}{\cos \frac{2\pi }{9}}%
, \\
\tilde{\chi}_{\lambda _{4}} &=&\frac{8}{3}\,\tilde{\chi}_{\lambda _{2}}\,%
\tilde{\chi}_{\lambda _{3}}\,\frac{\sin \frac{\pi }{18}\cos \frac{\pi }{9}}{%
\sin \frac{7\pi }{18}},\,\,\,\,\,\,\,\,\,\,\tilde{\chi}_{2\lambda _{1}}=4\,%
\tilde{\chi}_{\lambda _{1}}\,\cos \frac{2\pi }{9}\cos \frac{\pi }{9}%
,\,\,\,\,\,\,\,\,\,\,\tilde{\chi}_{2\lambda _{2}}=2\tilde{\chi}_{2\lambda
_{1}} \\
\tilde{\chi}_{2\lambda _{3}} &=&2\sqrt{3}\,\tilde{\chi}_{3\lambda _{1}}\,%
\frac{\cos \frac{2\pi }{9}}{\sin \frac{5\pi }{18}},\,\,\,\,\,\,\,\,\,\,%
\tilde{\chi}_{2\lambda _{4}}=36\,\tilde{\chi}_{3\lambda _{2}}^{2}\,\cos ^{2}%
\frac{2\pi }{9},\,\,\,\,\,\,\,\,\,\,\tilde{\chi}_{3\lambda _{1}}=2\,\frac{%
\sin \frac{4\pi }{9}\sin \frac{7\pi }{18}}{\sin \frac{\pi }{18}\sin \frac{%
\pi }{9}},\,\,\quad  \\
\tilde{\chi}_{3\lambda _{2}} &=&\tilde{\chi}_{3\lambda _{3}}=\tilde{\chi}%
_{3\lambda _{1}}+2=\frac{\sin \frac{5\pi }{18}\sin \frac{7\pi }{18}}{\sin 
\frac{4\pi }{9}\sin \frac{\pi }{9}},\,\,\,\,\,\,\,\,\,\,\tilde{\chi}%
_{3\lambda _{4}}=\tilde{\chi}_{2\lambda _{4}},
\end{eqnarray}
such that we find the following relations amongst them 
\begin{eqnarray}
\tilde{\chi}_{\lambda _{1}} &=&2(1-\chi_{2\lambda _{1}}-\chi _{3\lambda
_{1}}-\chi _{6\lambda _{2}}) +\chi _{4\lambda _{1}}+\chi _{5\lambda _{2}}, \\
\tilde{\chi}_{\lambda _{2}} &=&2(\chi _{5\lambda _{1}}-\chi _{2\lambda
_{1}}-\chi _{2\lambda _{2}}-\chi _{6\lambda _{2}})+\chi _{5\lambda _{2}}, \\
\tilde{\chi}_{\lambda _{3}} &=&2(\chi _{\lambda _{1}}-\chi _{2\lambda
_{1}}+\chi _{5\lambda _{1}}-\chi _{2\lambda _{2}}-\chi _{3\lambda _{2}}+\chi
_{5\lambda _{2}}-\chi _{6\lambda _{2}}+1), \\
\tilde{\chi}_{\lambda _{4}} &=&2(1-\chi _{\lambda _{1}}-\chi _{2\lambda
_{1}}-\chi _{3\lambda _{1}}+\chi _{4\lambda _{1}}+\chi _{5\lambda _{1}}+\chi
_{\lambda _{2}}-\chi _{2\lambda _{2}}+\chi _{3\lambda _{2}}-\chi _{6\lambda
_{2}}),\quad \\
\tilde{\chi}_{2\lambda _{1}} &=&2(\chi _{5\lambda _{1}}-\chi _{\lambda
_{1}}-\chi _{2\lambda _{1}}-\chi _{3\lambda _{1}})+\chi _{4\lambda
_{1}}-\chi _{2\lambda _{2}}, \\
\tilde{\chi}_{2\lambda _{3}} &=&2(\chi _{5\lambda _{1}}-\chi _{\lambda
_{1}}-\chi _{2\lambda _{1}}-\chi _{3\lambda _{1}})-\chi _{4\lambda
_{1}}+\chi _{2\lambda _{2}}, \\
\tilde{\chi}_{2\lambda _{4}} &=&2(1-\chi _{\lambda _{1}}-\chi _{2\lambda
_{1}})-\chi _{3\lambda _{1}}+\chi _{4\lambda _{1}}+\chi _{6\lambda _{1}}, \\
\tilde{\chi}_{3\lambda _{1}} &=&2(\chi _{4\lambda _{1}}+\chi _{6\lambda
_{1}}-\chi _{\lambda _{1}}-\chi _{2\lambda _{1}}-\chi _{3\lambda _{1}}-\chi
_{5\lambda _{1}}), \\
\tilde{\chi}_{3\lambda _{2}} &=&\tilde{\chi}_{3\lambda _{3}}=2+\tilde{\chi}%
_{3\lambda _{1}}\,.
\end{eqnarray}
The recurrence relations (\ref{q1})-(\ref{q3}) are then solved by 
\begin{eqnarray}
Q_{1}^{1} &=&2{\chi }_{\lambda _{2}}-{\chi }_{\lambda _{1}}-{\chi }%
_{2\lambda _{1}}, \\
Q_{1}^{2} &=&{\chi }_{5\lambda _{1}}+{\chi }_{\lambda _{2}}+{\chi }%
_{6\lambda _{2}}-{\chi }_{\lambda _{1}}-{\chi }_{2\lambda _{1}}-{\chi }%
_{3\lambda _{1}}-{\chi }_{3\lambda _{2}}, \\
Q_{1}^{3} &=&{\chi }_{\lambda _{1}}+{\chi }_{2\lambda _{1}}+{\chi }%
_{6\lambda _{1}}+{\chi }_{4\lambda _{2}}-{\chi }_{5\lambda _{2}}-{1,} \\
Q_{1}^{4} &=&1-{\chi }_{4\lambda _{2}}-2({\chi }_{\lambda _{1}}+{\chi }%
_{2\lambda _{1}}+{\chi }_{3\lambda _{1}}+{\chi }_{4\lambda _{1}}+{\chi }%
_{5\lambda _{1}}-{\chi }_{6\lambda _{1}}+{\chi }_{2\lambda _{2}}-{\chi }%
_{5\lambda _{2}}), \\
Q_{2}^{1} &=&1+{\chi }_{\lambda _{1}}+{\chi }_{2\lambda _{1}}+{\chi }%
_{4\lambda _{1}}-{\chi }_{5\lambda _{1}}+{\chi }_{2\lambda _{2}}+{\chi }%
_{4\lambda _{2}}-{\chi }_{5\lambda _{2}}, \\
Q_{2}^{2} &=&{\chi }_{\lambda _{1}}+{\chi }_{2\lambda _{1}}-{\chi }%
_{3\lambda _{1}}-{\chi }_{4\lambda _{1}}+{\chi }_{5\lambda _{1}}+{\chi }%
_{6\lambda _{1}}+{\chi }_{\lambda _{2}}+{\chi }_{3\lambda _{2}}-{\chi }%
_{4\lambda _{2}}+{\chi }_{6\lambda _{2}}-1,\quad  \\
Q_{2}^{3} &=&2({\chi }_{3\lambda _{2}}+{\chi }_{5\lambda _{2}}-{\chi }%
_{\lambda _{1}}-{\chi }_{2\lambda _{1}}-{\chi }_{3\lambda _{1}}-{\chi }%
_{4\lambda _{1}}-{\chi }_{6\lambda _{2}})-{\chi }_{5\lambda _{1}}-{\chi }%
_{6\lambda _{1}}-{\chi }_{2\lambda _{2},} \\
Q_{2}^{4} &=&8({\chi }_{3\lambda _{2}}+{\chi }_{4\lambda _{2}}+{\chi }%
_{5\lambda _{2}}-{\chi }_{\lambda _{1}}-{\chi }_{2\lambda _{1}}-{\chi }%
_{3\lambda _{1}}-{\chi }_{4\lambda _{1}}-{\chi }_{5\lambda _{1}}-{\chi }%
_{\lambda _{2}}-1)  \nonumber \\
&&-7{\chi }_{6\lambda _{1}}+6{\chi }_{6\lambda _{1}}.
\end{eqnarray}
Using these values we compute numerically the central charge to $c=16$.

\section{Unstable quasi-particles}

Once a character can be expressed in the generic form (\ref{chi}), it does
not only allow a derivation of the constant TBA equations, but also, when
interpreted as partition function, one may construct quasi-particle spectra
of different statistical nature. We proceed in the usual fashion, but we
will now introduce as the main novelty also unstable quasi-particles inside
the spectrum. As usual \cite{KM} we parameterize the partition function $%
\chi (q=e^{2\pi v/ktL})$ by Boltzmann's constant$\ k$, the temperature $T$,
the size of the quantizing system $L$, and the speed of sound $v$. We then
equate it with $\sum_{n=0}^{\infty }P(E_{n})\exp (-E_{n}/kT)$, where $%
P(E_{n})$ denotes the degeneracy of the particular energy level $%
E_{n}=E_{n}(p_{A})$ as a function of the single particle contributions of
type $A$. It is the aim in this analysis to identify the spectrum expressed
in terms of the $p_{A}$. Technically this can be achieved by making use of
the expressions for the number of partitions $\mathcal{Q}_{s}(n,m)$ ($%
\mathcal{P}_{s}(n,m)$) of the positive integer $n$ into $m$ non-negative
(distinct) integers smaller or equal to $s$ (see e.g. \cite{Andr}) 
\begin{equation}
\sum_{n=0}^{\infty }\mathcal{P}_{s}(n,m)q^{n}=q^{m(m-1)}\left[ \QATOPD. .
{s+1}{m}\right] _{q},\quad \quad \sum_{n=0}^{\infty }\mathcal{Q}%
_{s}(n,m)q^{n}=\left[ \QATOPD. . {s+m}{m}\right] _{q}\,\,.
\end{equation}
Introducing in the standard way \cite{KM} some internal quantum numbers we
construct for instance (in units of $2\pi /L$) a purely fermionic 
\begin{equation}
p_{N_{a}}^{a}(\vec{k})=\frac{1}{2}([M_{ab}]_{q}-\delta _{ab})k_{b}+\frac{1}{2%
}+B_{a}+N_{a}
\end{equation}
or purely bosonic 
\begin{equation}
p_{N_{a}}^{a}(\vec{k})=\frac{1}{2}[M_{ab}]_{q}k_{b}+B_{a}+\hat{N}_{a}
\end{equation}
quasi-particle spectrum. The positive integers $N_{a}$ and $\hat{N}_{a}$ are
constrained from above as $N_{a}<$Int$((1-[M_{ab}]_{q})$ $%
k_{b}+B_{a}^{\prime }$ $)$ and $\hat{N}_{a}\leq $Int$((1-[M_{ab}]_{q})$ $%
k_{b}+m_{a}+B_{a}^{\prime })$, with Int$(x)$ to be the integer part of $x$.
Like in the non-deformed case, it is of course also possible to construct
spectra related to more exotic or even with mixed statistics.

We expect now that at a certain energy scale some unstable particles vanish
from the spectrum. The mechanisms for this is that the upper bounds $N_{a},%
\hat{N}_{a}$ involved in the expressions for the possible momenta $%
p_{N_{a}}^{a}(\vec{k}),$ $p_{\hat{N}_{a}}^{a}(\vec{k})$ decrease. We
illustrate this with some examples. Denoting the character for the vacuum
sector of the minimal model $\mathcal{M}(k,k+1)$ by $\chi ^{k}(q)$ \cite
{minchar}, we compute for instance 
\begin{eqnarray}
\chi ^{2}(q)-\chi ^{1}(q) &=&{q^{6}}+{q^{7}}+2\,{q^{8}}+3\,{q^{9}}+5\,{q^{10}%
}+6\,{q^{11}}+9\,{q^{12}}+11\,{q^{13}}+16\,{q^{14}}+  \nonumber \\
&&20\,{q^{15}}+27\,{q^{16}}+33\,{q^{17}}+44\,{q^{18}}+54\,{q^{19}}+70\,{%
q^{20}}+\mathcal{O}(q^{21}),
\end{eqnarray}

This means for example comparing $\chi ^{1}(q)$ and $\chi ^{2}(q)$ one
particle should vanish from the spectrum of $\mathcal{M}(2,3)$ at level 6
when we vary the value of the resonance parameter such that it flows to $%
\mathcal{M}(1,2)$. Indeed in the purely fermionic spectrum we have the
possibility of a six particle contribution involving four of type 1 and two
particles of type 2 with $N_{2}<$Int$(2[(1-\exp (-r/2m_{2}))+\exp
(-r/2e^{|\sigma _{12}|/2})])$. This means for $rm_{2}/2\ll 1$ and $%
r/2e^{|\sigma _{12}|/2}\ll 1$ the state 
\begin{equation}
\left|
p_{0}^{1}(4,2),p_{1}^{1}(4,2),p_{2}^{1}(4,2),p_{3}^{1}(4,2),p_{0}^{2}(4,2),p_{1}^{2}(4,2)\right\rangle \quad
\end{equation}
is allowed. It is then clear that when we increase $\sigma _{12}$, this
state disappears from the spectrum. At the same time the state 
\begin{equation}
\left|
p_{0}^{1}(4,2),p_{1}^{1}(4,2),p_{2}^{1}(4,2),p_{4}^{1}(4,2),p_{0}^{2}(4,2),p_{1}^{2}(4,2)\right\rangle \quad
\end{equation}
at level 7 and the two states 
\begin{eqnarray}
&&\left|
p_{0}^{1}(4,2),p_{1}^{1}(4,2),p_{2}^{1}(4,2),p_{5}^{1}(4,2),p_{0}^{2}(4,2),p_{1}^{2}(4,2)\right\rangle
\\
&&\left|
p_{0}^{1}(4,2),p_{1}^{1}(4,2),p_{3}^{1}(4,2),p_{4}^{1}(4,2),p_{0}^{2}(4,2),p_{1}^{2}(4,2)\right\rangle
\end{eqnarray}
at level 8, etc. vanish for the same reason.

\section{Conclusions}

We have demonstrated that it is possible to construct scaling functions
which reproduce the renormalization group flow by q-deforming fermionic
versions of Virasoro characters in a very natural way. We investigated a
fairly generic class of theories related to a pair of simple simply laced
Lie algebras \textbf{g} and \textbf{\~{g} }or associated coset models. The
construction procedure relies on the fact that the characters, quantities of
the massless theory, involve data of the massive theory, i.e. the phases of
the S-matrices. At the fixed points of these flows we solved the relevant
recurrence relations analytically in terms of Weyl characters. We provided
here various new solutions for particular choices of the algebras involved.
It would be extremely interesting to answer the question whether it is
possible to solve these relations in a completely generic, i.e.
case-independent fashion. One should note that our solutions admit various
ambiguities, i.e. the sums are not unique since there are numerous character
identities involved or they might be expressed in terms of direct products
of characters in a Clebsch-Gordan sense. This arbitrariness might be
eliminated when one possible finds a deeper interpretation of the recurrence
relation in terms of representation theory.

Furthermore, it would be interesting to investigate whether it is possible
to modify the Weyl characters, for instance by a specific choice of the $%
\tau $'s, in such a way that they solve the full $r$-dependent recurrence
relations (\ref{Qr}) exactly. Noting that our scaling functions only
coincide qualitatively with those obtained from the full TBA analysis, in
the sense that they have the plateaux precisely in the same position,
including their size in the $r$-direction, one may ask a stronger question:
Is it possible to find versions of Weyl characters such that the full TBA
equations, this would be their formulation in terms of so-called Y-systems
(see e.g. \cite{TBAZamun}), is reproduced?

The functions we constructed allow for a far easier investigation of the
RG-behaviour than the full TBA-system \cite{TBA}, the scaled c-theorem \cite
{ZamC,CF2} or the semi-classical analysis \cite{Zamref}. This allows to
investigate systems of more complex nature such as $A_{1}|E_{6}$ or flows
between different supersymmetric series. It would be interesting to
investigate the latter flow in the other approaches.

The level-rank duality of the type (\ref{HSGmin}) gives a hint why it is
possible to obtain the same flow by means of a theory involving unstable
particles and alternatively as massless flows in the sense of \cite{Stair}.
The concrete link, however, i.e. the question of how this duality is
reflected in the massive models, that is the scattering matrix, is still
eluded from our analysis.

We have also shown that our q-deformed characters allow for the construction
of spectra, which involve also unstable quasi-particles. The ``decay'' of
these particles from the spectrum is governed by a variable bound on the
momenta depending on the resonance parameter.

Concerning the specific theories investigated, it would be of interest to
extend the analysis to models which involve also non-simply laced algebras,
albeit for \textbf{g} non-simply laced consistent S-matrices have not been
constructed at present.\medskip

\newpage

\noindent \textbf{Acknowledgments: } We are grateful to the Deutsche
Forschungsgemeinschaft (Sfb288), PGIDT-PXI-2069, CICYT (AEN99-0589) and
DGICYT (PB96-0960) for financial support. A.F. thanks the Departamento de
F\'{\i}sica de Part\'{\i}culas of the Universidade de Santiago de
Compostela, where part of this work was carried out, for their kind
hospitality.

\end{document}